\documentclass[footinbib,a4paper,aps,superscriptaddress,reprint,twocolumn,preprintnumbers,amsmath,amssymb,nobalancelastpage,10pt]{revtex4-1}
\usepackage{color} 
\usepackage{bm}
\usepackage{dcolumn}
\usepackage{times}
\usepackage{amssymb} 
\usepackage{amsmath} 
\usepackage{ragged2e}
\usepackage{graphicx}
\usepackage{epstopdf}
\usepackage[english]{babel}
\usepackage{mathrsfs}
\usepackage{textcomp}
\usepackage{amsthm}
\usepackage{amsfonts}
\usepackage[nice]{nicefrac}
\usepackage{soul}
\usepackage{braket} 
\usepackage[colorlinks=true,citecolor=blue,linkcolor=blue,urlcolor=blue]{hyperref}
\usepackage{cleveref}
\newcommand{\be}{\begin{equation}}
\newcommand{\ee}{\end{equation}}
\newcommand{\bea}{\begin{eqnarray}}
\newcommand{\eea}{\end{eqnarray}}
\newcommand{\ba}{\begin{array}}
\newcommand{\ea}{\end{array}}

\begin{document}

\title{Higher-order topological Peierls insulator in a two-dimensional atom-cavity system}

\author{Joana Fraxanet}
\email{jfraxanet@icfo.eu}
\affiliation{ICFO - Institut de Ci\`encies Fot\`oniques, The Barcelona Institute of Science and Technology, 08860 Castelldefels (Barcelona), Spain}

\author{Alexandre Dauphin}
\affiliation{ICFO - Institut de Ci\`encies Fot\`oniques, The Barcelona Institute of Science and Technology, 08860 Castelldefels (Barcelona), Spain}

\author{Maciej Lewenstein}
\affiliation{ICFO - Institut de Ci\`encies Fot\`oniques, The Barcelona Institute of Science and Technology, 08860 Castelldefels (Barcelona), Spain}
\affiliation{ICREA, Pg. Lluis Companys 23, ES-08010 Barcelona, Spain}

\author{Luca Barbiero}
\affiliation{Institute for Condensed Matter Physics and Complex Systems,
DISAT, Politecnico di Torino, I-10129 Torino, Italy}

\author{Daniel Gonz\'alez-Cuadra}
\email{daniel.gonzalez-cuadra@uibk.ac.at}
\affiliation{Institute for Theoretical Physics, University of Innsbruck, 6020 Innsbruck, Austria}
\affiliation{Institute for Quantum Optics and Quantum Information of the Austrian Academy of Sciences,
6020 Innsbruck, Austria}

\date{\today}

\begin{abstract}
In this work, we investigate a two-dimensional system of ultracold bosonic atoms inside an optical cavity, and show how photon-mediated interactions give rise to a plaquette-ordered bond pattern in the atomic ground state. The latter corresponds to a 2D Peierls transition, generalizing the spontaneous bond dimmerization driven by phonon-electron interactions in the 1D Su-Schrieffer-Heeger (SSH) model. Here the bosonic nature of the atoms plays a crucial role to generate the phase, as similar generalizations with fermionic matter do not lead to a plaquette structure. Similar to the SSH model, we show how this pattern opens a non-trivial topological gap in 2D, resulting in a higher-order topological phase hosting corner states, that we characterize by means of a many-body topological invariant and through its entanglement structure. Finally, we demonstrate how this higher-order topological Peierls insulator can be readily prepared in atomic experiments through adiabatic protocols. Our work thus shows how atomic quantum simulators can be harnessed to investigate novel strongly-correlated topological phenomena beyond those observed in natural materials.
\end{abstract}


\maketitle

\paragraph*{Introduction.--}

In the last decades, topology has reached a central role in the description and classification of phases of matter~\cite{Moore_2010}. Contrary to the standard Landau paradigm~\cite{Landau2937}, topological phases are not distinguished by different spontaneous symmetry-breaking (SSB) patterns, but rather by different non-local topological invariants~\cite{Hasan_2010, Qi_2011}. Non-trivial topology manifests itself in the presence of conducting edge states in bulk insulating phases~\cite{MacDonald_1990, Hatsugai_1993, Ryu_2002}, quantized conductances~\cite{von,laughlin,thouless,moore} or fractional charges~\cite{tsui,laughlin1,su}, which in the case of symmetry-protected topological (SPT) phases are robust against perturbations that do not break certain protecting symmetries~\cite{chiu2016}. Recently, this class has been enlarged to include higher-order SPT (HOSPT) phases~\cite{benalcazar2017,Benalcaza_2017_2}, protected by crystalline symmetries and hosting edge states of co-dimension larger than one, such as corner states in 2D~\cite{schindler2018, Khalaf_2018}. Despite the recent progress in the study of interacting HOSPT phases~\cite{You_2018, Dubinkin_2019, Kudo_2019, Laubscher_2019, Laubscher_2020, Sil_2020, Rasmussen_2020, bibo2020, Peng_2021, Guo_2022, Hackenbroich_2021, Otsuka_2021, gonzalez2022, Li_2022, Montorsi_2022, Wienand_2022, Aksenov_2023}, a full classification is still lacking.

One of the earliest examples of a SPT phase is found in the Su-Schrieffer-Heeger (SSH) model for polyacetylene~\cite{ssh}. In this 1D chain, interactions between electrons and phonons induces a so-called Peierls instability~\cite{Peierls_1955}, giving rise to a $\mathbb{Z}_2$ SSB characterized by a dimerized bond pattern, resulting in a topologically non-trivial gapped bulk protected by a chiral symmetry~\cite{Asboth_2016}. In this case, the interplay between symmetry breaking and symmetry protection gives rise, for instance, to delocalized fractionally-charged quasi-particles~\cite{Heeger_1988} and topological quantum critical points~\cite{fraxanet}, absent in the non-interacting case~\cite{Hasan_2010, Qi_2011}. 

Generalizations of the SSH model to 2D do not result, however, in non-trivial (higher-order) topology phases~\cite{Xing2021} As it was first shown by Benalcazar, Bernevig and Hughes (BBH)~\cite{benalcazar2017, Benalcaza_2017_2}, a 2D fermionic system with dimerized bonds in both the $x$ and $y$ direction (forming a plaquette structure), can host a HOSPT phase, but only in the presence of a non-zero flux that additionally breaks time-reversal symmetry. This symmetry-broken pattern is not achieved spontaneously starting from a symmetric fermion-phonon model, as shown by recent Monte Carlo studies~\cite{Xing2021}. 

The situation is different, however, for the bosonic BBH model~\cite{Dubinkin_2019, bibo2020}, where a topologically non-trivial bulk gap appears without the SSB of time reversal. A natural question is therefore whether the corresponding HOSPT phase can emerge through a SSB process from a $\mathbb{Z}_2\times \mathbb{Z}_2$ symmetric bosonic Hamiltonian, where, similarly to the 1D case~\cite{chanda2021}, the interplay between symmetry breaking and symmetry protection could be further explored. Contrary to solid-state materials, which possess fermionic particles as fundamental constituents, ultracold atoms in optical lattices offer the unique opportunity to investigate strongly-correlated bosonic matter in highly-controllable experimental conditions~\cite{Jaksch_2005, Lewenstein_2007, Gross_2017}. Although optical lattices are static, the role of phonons can be emulated using a second atomic species~\cite{gonzalez2018} or by placing the atomic gas inside a cavity~\cite{Ritsch_2013}, where photons mediate infinite-range interactions between atoms. In the 1D case, these synthetic phonons indeed drive similar solid-state-like phenomena, including topological Peierls insulators~\cite{Mivehvar_2017, gonzalez2019, gonzalez2019b,chanda2021, chanda2022, Chanda_2022, manmana,sirker,wang,barbiero2,sbierski,yu} and fractionally-charged quasi-particles~\cite{Fraser_2019, Gonzalez-Cuadra_2020a, Gonzalez-Cuadra_2020b}, also predicted in the presence of dipolar interactions~\cite{Sergi2022}.

In this work, we extend these results to 2D by considering a system of ultracold bosonic atoms coupled to a cavity, a setup that has been previously employed to prepare strongly-correlated phases such as supersolids or charge density waves~\cite{Esslinger2016, Leonard_2017}. Here we show how, by simply modifying the phase of the cavity mode, a plaquette-ordered structure with dimerization both in the $x$ and $y$ directions appears, driven in this case by photon-mediated interactions. This symmetry-broken pattern is indeed the same as the one required in the bosonic BBH model, but here it emerges spontaneously through a bosonic Peierls transition, instead of being imposed externally. Contrary to the 1D case~\cite{chanda2021}, where fermions and (hardcore) bosons are esentially equivalent, our results show how bosonic particles in 2D can self-organize in an intrinsically new phase of matter, that cannot be generated with a symmetric fermion-phonon model~\cite{Xing2021} We confirm moreover the topological nature of this phase, that we denote higher-order topological Peierls insulator (HOTPI), using a many-body topological invariant as well as through its entanglement structure. Finally, we propose and benchmark a quantum simulation protocol to adiabatically prepare this phase in current atomic experiments.

\begin{figure}
\includegraphics[width=0.95\columnwidth]{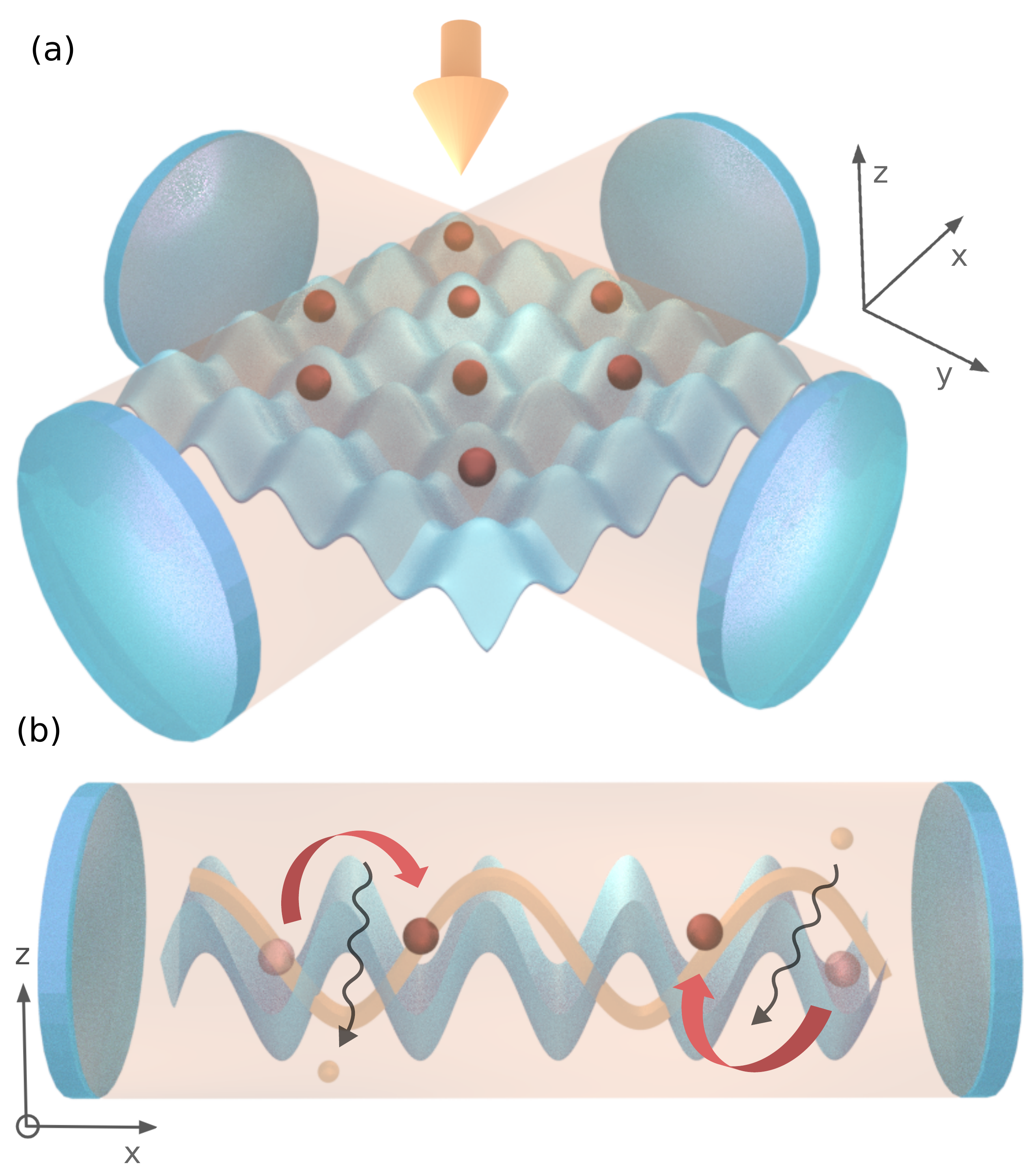}
\caption{\label{fig:fig0} \textbf{Atom-cavity experimental setup:}
(a) Ultracold bosonic atoms are trapped in the lowest band of a 2D optical lattice. The atoms are coupled to two cavity modes created by two optical cavities aligned in the $x$ and $y$ directions, and to a laser pump aligned in the $z$ direction. (b) In each direction, the relative phase between the optical lattice (blue) and the cavity mode (yellow) is chosen such that the nodes of the latter coincide with the sites of the lattice. In this configuration, the effective Hamiltonian describing the atom-cavity system contains correlated-tunneling terms, where atoms can tunnel between n.n. sites by absorbing or emitting a photon from the cavity.}
\end{figure}

\vspace{0.1cm}

\paragraph*{Bose-Hubbard model with cavity-mediated correlated tunneling.--}
We consider a system of ultracold bosonic atoms trapped in the lowest band of a $L\times L$ square optical lattice. The atoms are additionally coupled to two cavity modes created by two optical cavities aligned along the $x$ and $y$ directions, together with a laser pump in the $z$ direction [Fig.~\ref{fig:fig0}(a)].  In the basis of localized Wannier functions, the system is described by the following  Hamiltonian~\cite{maschler2005},
\begin{equation}
\label{eq:H_mode}
    \begin{aligned}
    H =& -t\sum_{\langle i,j \rangle} \left(b_i^\dagger b^{\vphantom{\dagger}}_j + \text{H.c.}\right) + \frac{U}{2}\sum_j n_j (n_j -1) \\
    & - \Delta_c \sum_\mu a_\mu^\dagger a^{\vphantom{\dagger}}_\mu +  g\sum_\mu(a^{\vphantom{\dagger}}_\mu+a_\mu^\dagger) B_\mu,
    \end{aligned}
\end{equation}
with $B_\mu = \sum_{i} (-1)^{i+1} \left(b_i^\dagger b^{\vphantom{\dagger}}_{i+\hat{\mu}} + \text{H.c.}\right)$. Here  $b^\dagger_i/b^{\vphantom{\dagger}}_i$ ($a^\dagger_\mu/a^{\vphantom{\dagger}}_\mu$) denotes the bosonic creation/annihilation operators of an atom (photon) at site $i$ (mode $\mu\in \{x,y\}$). The parameters $t$ and $U$ correspond to the standard nearest-neighbor (n.n.) tunneling and Hubbard onsite interactions~\cite{Jaksch_1998}, respectively, and $\Delta_c$ is the cavity-pump detuning~\cite{maschler2005}. Finally, $g$ characterizes the interactions between the atoms and the cavity, where, differently to previous experiments~\cite{Esslinger2016, Leonard_2017}, the latter is not coupled to the atomic density but to the staggered tunneling $B_\mu$. Similarly to the 1D case~\cite{chanda2021}, this is achieved by considering a relative phase $\phi = \pi / 2$ between the cavity modes and the optical lattice [Fig.~\ref{fig:fig0}(b)] (see~\cite{SM} for a full derivation).

The atom-cavity Hamiltonian is invariant under $\mathbb{Z}_2\times\mathbb{Z}_2$ transformations, where each $\mathbb{Z}_2$ symmetry corresponds to a one-site translation of the atoms in the direction $\mu$ together with the transformation $a_\mu \rightarrow -a_\mu$. We anticipate that the HOTPI phase will be driven by the spontaneous breaking of this symmetry. Before analyzing the phase diagram of the model, we first obtain an effective atomic Hamiltonian by adiabatically eliminating the cavity modes in the presence of a large cavity decay rate~\cite{maschler2005}, resulting in the limit of large detuning $\Delta_c$ in the following expression (see~\cite{SM}),
\begin{equation}\label{eq:H_tb}
     H_{\rm eff} = -t \sum_{\langle i, j\rangle} (b_i^\dagger b^{\vphantom{\dagger}}_{j} + \text{H.c.}) + \frac{U}{2} \sum_{i=j}^{N} n_j(n_j -1) + \frac{U_C}{L^2}\sum_\mu B_\mu^2,
\end{equation}
with $U_C = g^2L^2/\Delta_C$. The last term in Eq.~\eqref{eq:H_tb} accounts for the cavity-mediated all-to-all interactions arising from the correlated tunneling terms in Eq.~\eqref{eq:H_mode}. The latter differs from the cavity-mediated density-density interaction considered in Ref.~\cite{Esslinger2016, Leonard_2017} for the case $\phi = 0$, giving rise to a charge density wave and a supersolid phase. As we show below, a HOTPI phase can be obtained instead with the same experimental setup just by modifying the relative phase to $\phi = \pi/2$.

%
%
%
%
%
%
\begin{figure}
\includegraphics[width=1\columnwidth]{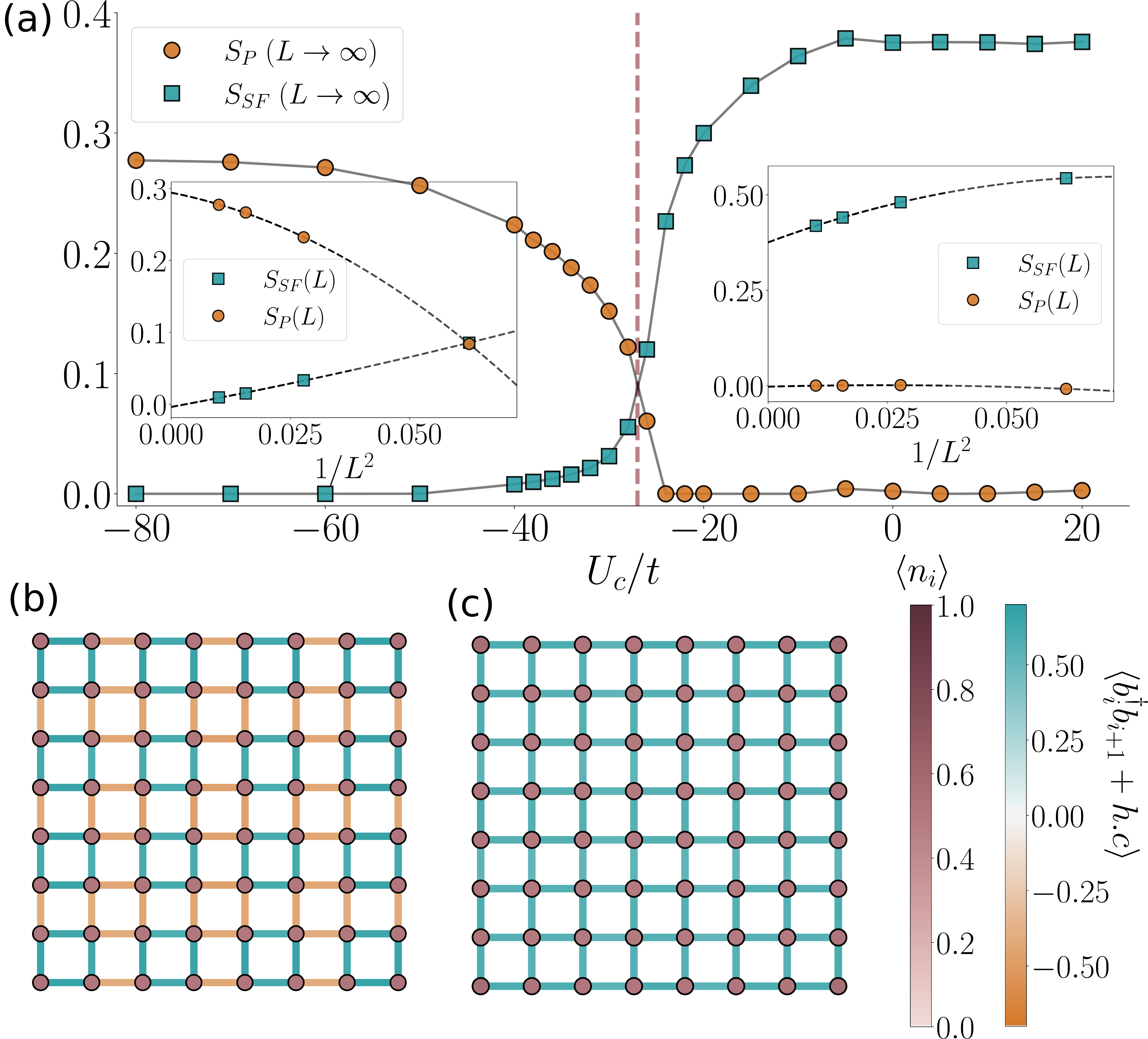}
\caption{\label{fig:fig1} \textbf{Phase diagram:} (a) SF ($S_{\rm SF}$) and plaquette ($S_{\rm P}$) order parameters calculated by a finite-size scaling to the thermodynamic limit, indicating the presence of a plaquette-ordered phase for $U_c/t \lessapprox -27$, and a SF phase for $U_c/t \gtrapprox -27$.  The insets show the finite-size scaling at $U_c/t = -80$ (left) and $U_c/t = 10$ (right) for system sizes $L\times L$, with $L\in\{4,6,8,10\}$. (b) and (c) show the real-space configuration of the plaquette-ordered phase at $U_c/t=-80$ and the SF phase at $U_c/t = 30$, respectively. The color of the lattice sites and bonds denote the expectation value of the onsite occupation $\langle n_i\rangle$ and the n.n. tunneling $\langle b_i^\dagger b^{\vphantom{\dagger}}_j + \text{H.c.} \rangle$, respectively. 
}
\end{figure}
%
%


\vspace{0.1cm}

\paragraph*{Bosonic Peierls transition in 2D.--} We study the phase diagram of the effective Hamiltonian \eqref{eq:H_tb} in the limit of hardcore bosons ($U\gg t, U_C$) as a function of the strength $U_c$ of the cavity-induced interactions. In the following, and unless stated otherwise, we fix the atomic density to half filling, $N = L^2/2$, and consider open boundary conditions. The ground state of the system is calculated through a density-matrix renormalization group (DMRG) algorithm~\cite{dmrg} based on matrix product states~\cite{tenpy}, where we consider bond dimensions up to $\chi = 1000$.


As we show in Fig.~\ref{fig:fig1}(a), for positive or small negative values of $U_c$, the system is in a superfluid (SF) phase with off-diagonal long-range order, characterized by a finite value of the single-particle momentum distribution $S(\mathbf{q})$ at momentum $\mathbf{q} = (0,0)$, $S_{\rm SF} \equiv S(0)$, with
\begin{equation}\label{eq:S0}
    S(\vec{q}) = \frac{1}{L^2} \sum_{i,j} \langle b^\dagger_i b^{\vphantom{\dagger}}_j + \text{H.c.} \rangle e^{i\mathbf{q}(\mathbf{r_i}-\mathbf{r}_j)},
\end{equation}
where the indices in the sum run over every pair of sites. For $U_c/t \lessapprox  -27$, on the other hand, the cavity-mediated interactions induce a plaquette-ordered phase~\cite{shoushu2014,wangling2016} through a spontaneous breaking of the $\mathbb{Z}_2\times\mathbb{Z}_2$ symmetry. The latter is characterized by a finite value of the plaquette structure factor~\cite{mambrini2006},
\begin{eqnarray}\label{eq:SPVBS}
\begin{aligned}
    S_{\rm P} &= \frac{1}{(L-1)L -3} \sum_{\langle k,l\rangle} \epsilon(k,l) \\ 
    & \times\left[ \langle b^\dagger_i b^{\vphantom{\dagger}}_j b^\dagger_k b^{\vphantom{\dagger}}_l + \text{H.c.} \rangle - \langle (b^\dagger_i b^{\vphantom{\dagger}}_j + \text{H.c.})^2\rangle \right],
\end{aligned}
\end{eqnarray}
where the sum $\langle k,l\rangle$ runs over n.n. pairs, $\langle i,j\rangle$ is a fixed bond in the middle of the lattice, and the form factor $\epsilon(k,l)$ is defined as in Ref.~\cite{mambrini2006}.

Figure~\ref{fig:fig1}(b) shows the real-space configuration of the plaquette-ordered phase, where the expectation value of the n.n. bosonic tunneling reveals a bond dimmerization both in the $x$ and $y$ directions, contrary to the translational invariant SF case [Fig.~\ref{fig:fig1}(c)]. Our results are summarized in Fig.~\ref{fig:fig1}(a), where the ground-state values of the order parameters $S_{\rm SF}$ and $S_{\rm P}$ are depicted for different system sizes. We perform a finite-size scaling for the latter deep in each phase, showing how $S_{\rm P}$ ($S_{\rm SF})$ goes to zero (finite value) for the SF phase in the thermodynamic limit ($L\rightarrow\infty)$, and viceversa for the plaquette-ordered phase. Finally, we show the extrapolated value of the order parameters as a function of $U_C/t$. In both cases, the results are consistent with a phase transition between a SF and a plaquette-ordered phase at $U_C/t\approx -27$. 

We note that the plaquette-ordered phase corresponds to a two-dimensional bosonic Peierls transition, where the $\mathbb{Z}_2\times\mathbb{Z}_2$ SSB is driven by atom-photon interactions. As we mentioned above, this is a generalization of the Peierls transition driven by electron-phonon interactions in the SSH model~\cite{ssh}. In this case, the bosonic nature of the atoms is crucial to obtain the plaquette-ordered structure, which is absent in similar generalizations of the SSH model to 2D, where only one $\mathbb{Z}_2$ symmetry is spontaneously broken~\cite{Xing2021}. Since this emergent plaquette structure corresponds to the one imposed externally in the bosonic BBH model~\cite{Dubinkin_2019, bibo2020}, it is therefore natural to ask whether the corresponding phase has a non-trivial topological nature.

\vspace{0.1cm}

\paragraph*{Higher-order topological Peierls insulator.--}
%
%

We analyze now the topological properties of the plaquette-ordered phase, showing that it corresponds to a HOSPT phase. Similarly to the case of the bosonic BBH model~\cite{Dubinkin_2019, bibo2020}, the latter is protected by a $U(1)\times C_4$ symmetry, where $U(1)$ corresponds to the particle number conservation and $C_4$ is the lattice rotational symmetry, preserved in this case by the plaquette structure after the $\mathbb{Z}_2\times\mathbb{Z}_2$ SSB. We confirm the topological nature of the phase through its entanglement spectrum structure~\cite{Pollmann2010} and through a many-body topological invariant~\cite{hatsugai2020}.

Let us start by studying the entanglement spectrum along different bipartitions. We first note that, as a result of the SSB, the plaquette-ordered ground state is four-fold degenerate, where the four configurations are connected by one-site translations along the $x$ and $y$ directions. More precisely, finite-size effects introduce energy splittings in the ground-state manifold, that vanishes exponentially by increasing the system size, and the exact degeneracy is recovered in the thermodynamic limit. Similarly to the SSH model, where the ground state is two-fold degenerate, only one configuration is topologically non-trivial~\cite{Asboth_2016}\textemdash in this case the one where the staggered value of $\langle b^\dagger_i b^{\vphantom{\dagger}}_j + \text{H.c}\rangle$ is smaller around the corners. It is important to stress that, although here the lowest-energy state for finite sizes is the one with stronger bonds around the corner [Fig.~\ref{fig:fig1}(b)], the topological configuration can be adiabatically prepared as a long-lived metastable state, as we show below.

\begin{figure}[t]
\includegraphics[width=1\columnwidth]{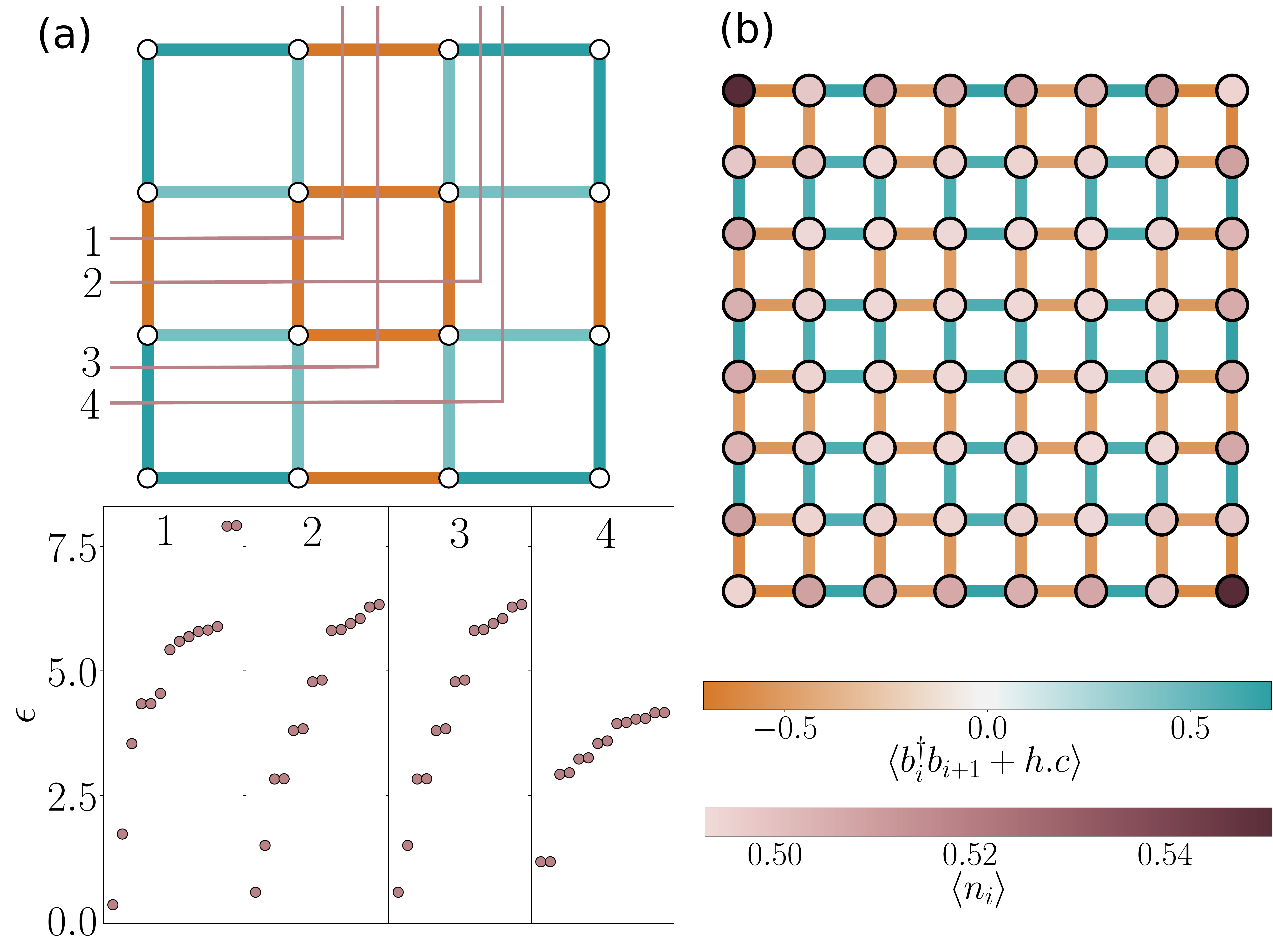}
\caption{\label{fig:fig2} \textbf{Entanglement spectrum structure and corner states.} (a) Lower eigenvalues of the entanglement spectrum for the bipartitions indicated in the figure ($1-4$). We consider the ground state of a system with size $L = 4$ at $U_c/t = -40$ (within the plaquette-ordered phase). (b) Real-space bond pattern and local occupation for the topological configuration at $U_c/t = -100$, for a system with $L = 10$. Note that this state is not the lowest energy state of the system due to the quasi-degeneracy of the ground state, but it is instead a metastable state. 
}
\end{figure}

Fig.~\ref{fig:fig2}(a) shows the entanglement spectrum for the ground state in the plaquette-ordered phase at different bipartitions. In general, the ground-state wavefunction can be written as $\vert \psi_{GS} \rangle = \sum_n e^{-\epsilon_n/2 } \vert \psi_n \rangle_A \otimes \vert \psi_n \rangle_B $, where $A$ and $B$ are two complementary subsets of lattice sites, and $\epsilon_n$ is the corresponding entanglement spectrum associated to that bipartition. For HOSPT phases, bipartitions that create virtual corners lead to even degenerate spectra~\cite{Pollmann2010}. In our case, and similarly to the bosonic BBH model, the broken $\mathbb{Z}_2\times\mathbb{Z}_2$ symmetry allows for four inequivalent bipartitions of this kind, and we observe how for one of them the spectrum is indeed degenerate [Fig.~\ref{fig:fig2}(a)]. The latter corresponds to the bipartition that creates a virtual corner surrounded by weak bonds, similar to the virtual edges in the 1D case~\cite{gonzalez2019}.

We note that the entanglement spectrum is a bulk quantity that provides information on the topological nature of the phase, but also indicates the possibility of finding localized states in a physical corner for finite systems. Although this is not a priori guaranteed by a bulk-boundary correspondence~\cite{Dubinkin_2019, bibo2020}, we found that this is indeed the case for the topological configuration of the plaquette-ordered phase, as depicted in Fig.~\ref{fig:fig2}(b). These corner states are not necessarily located in the middle of the gap, and their localization length can be quite large if they are close to bulk bands, as it appears to be the case here given the small extra atomic occupation with respect to half filling. A more in-depth characterization of these corner states requires therefore the study of larger system sizes.


%

We further confirm the topological nature of the plaquette-ordered phase using a many-body topological invariant. We compute in particular a many-body Berry phase $\gamma$ by locally deforming the Hamiltonian that introduces a flux through the central plaquette, as described in Ref.~\cite{hatsugai2020}, and then compute the discretized Wilson loop as we vary the flux following a close path $C$ in parameter space~\cite{Atala2013}. The Berry phase is then approximated by $\gamma_C^m = \text{Arg} \prod_{i=0}^{m-1} \langle \hat{\psi}_i \vert \hat{\psi}_{i+1}\rangle$, where $\ket{\psi_i}$ is the ground state for a given flux in the discretized loop of $m$ points. The Berry phase, independent on the path, is recovered in the limit of large $m$. We compute this quantity for a system with $L=10$ and obtain the values $\gamma_C = 4\cdot 10^{-5}$ and $\gamma_C=0.9995$ (mod $2 \pi$) for the SF and plaquette-ordered phases, respectively, confirming the topological nature of the latter.

\vspace{0.1cm}

\begin{figure}
\includegraphics[width=1\columnwidth]{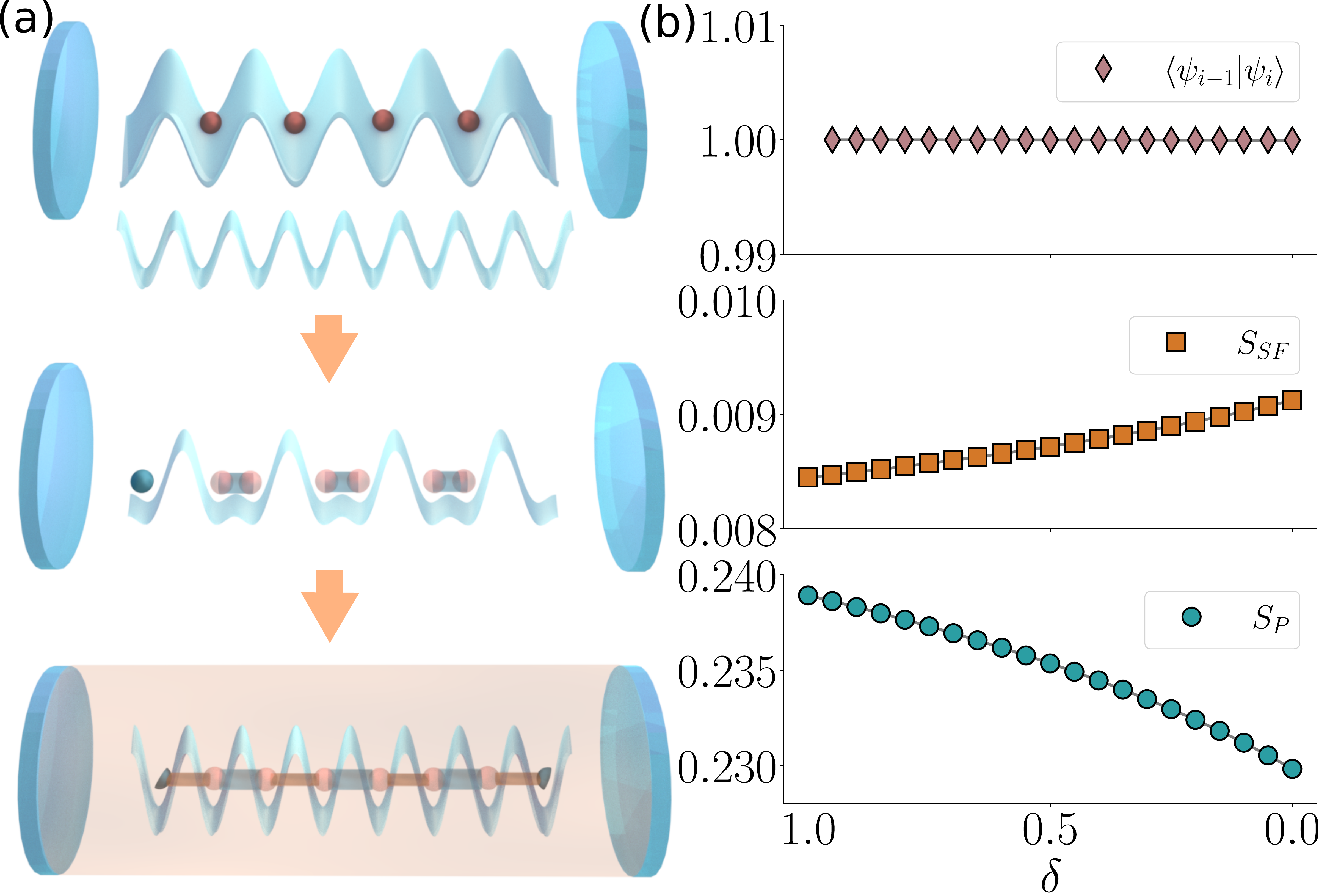}
\caption{\label{fig:fig3} \textbf{Adiabatic preparation of the HOTPI phase.} Sketch of the three main steps of the preparation scheme (a), starting with a deep lattice with wavelength $\lambda$, adding a second optical lattice with wavelength $\lambda/2$, and finally turning on the cavity and adiabatically eliminating the first lattice. (b) Quantum fidelity, SF and plaquette order parameters along the third step of the experimental protocol, where the two-dimensional dimerization $\delta$ is taken from $1$ to $0$ (see main text). The results are obtained for a system with $L^2=64$ and $U_c/t=-100$.}
\end{figure}

\paragraph*{Experimental realization.--}
We introduce now a protocol to prepare the HOTPI phase experimentally, which we depict in Fig.~\ref{fig:fig3}(a). In order to prepare the phase adiabatically, we need to break the $\mathbb{Z}_2\times\mathbb{Z}_2$ symmetry explicitly to avoid closing the gap at the Peierls transition, which moreover allows us to select the topological configuration among the quasi-degenerate ground-state manifold.  As a starting point, we consider a Mott insulator with unit filling in a square optical lattice with wavelength $\lambda$ in each direction, as routinely prepared in ultracold atomic experiments. Following Ref.~\cite{Lohse_2016}, we then introduce a second optical lattice with wavelength $\lambda/2$, creating a superlattice of double wells with weak bonds connecting the corners. The latter imprints the plaquette structure in the bosonic tunneling and breaks the $\mathbb{Z}_2\times\mathbb{Z}_2$ symmetry explicitly. At this point, the atomic Hamiltonian corresponds to the bosonic BBH model in the limit of zero inter-well tunneling~\cite{bibo2020}. Explicitly,
\begin{equation}
    H(\delta) = H_{\rm eff}(U_C=0) + \delta\sum_{i,\mu}(-1)^{s_i}\left(b^\dagger_i b^{\vphantom{\dagger}}_{i + \hat{\mu}} + \text{H.c.}\right)
\end{equation}
for the case positive $\delta = 1$, where $s_i = i_0 +i_1$ and $i = (i_0, i_1)$. We now turn on the cavity and tune the coupling $U_C$ to the desired value within the HOTPI phase. By slowly decreasing the intensity of the original $\lambda$-wavelength lattice we take $\delta$ to $0$, ending up with the atom-cavity Hamiltonian in Eq.~\ref{eq:H_tb}.

The protocol is adiabatic as the gap remains open along the whole path in parameter space. This is shown in Fig.~\ref{fig:fig3}(b), where we compute the ground-state fidelity after infinitesimal parameter changes along the path. We also show how, as expected, both the superfluid and the plaquette order parameters are consistent with a state in the HOTPI phase. Finally, we note that the topological configuration [Fig.~\ref{fig:fig2}(b)] can be selected simply by starting with the appropriate double-well structure. Even if for finite system sizes this configuration has a slightly higher energy than other configurations in the four-fold quasi-degenerate manifold, these are separated by energy barriers that diverge with the system size, guarantying the stability of the topological configuration. 

Regarding detection, we note that the plaquette structure can be directly observed with a quantum gas microscope~\cite{Bakr_2009, Sherson_2010}, while the non-trivial topological nature of the phase can be revealed by measuring the entanglement spectrum~\cite{Kokail_2021, Kokail_2021_2}.

\vspace{0.1cm}

\paragraph*{Conclusions and outlook.--} In this work, we showed how photon-mediated interactions in a two-dimensional bosonic system can give rise to a HOTPI phase. The latter can be prepared experimentally using ultracold bosonic atoms in optical lattices, by coupling the latter to two different cavity modes. We showed how the cavity-mediated all-to-all atomic interactions drive a Peierls transition, giving rise to a plaquette-ordered phase if the relative phases between the lattice and the cavity modes are properly chosen. Finally, we demonstrated the topological nature of the phase and proposed an adiabatic protocol to prepare it in current atomic experiments. In the future, it would be interesting to extend the setup and include multi-mode cavities, allowing to generate atom-photon topological defects that would generalize the topological solitons and fractionalized quasi-particles found in the SSH model to 2D~\cite{ssh}. Moreover, by exploring the regime of softcore bosons, we expect to find plaquette-ordered supersolid phases, similar to the density-ordered supersolids found in Ref.~\cite{Esslinger2016, Leonard_2017}.



\vspace{0.1cm}

\textbf{Acknowledgments}
We thank P. Stammer for discussion. ICFO group acknowledges support from: ERC AdG NOQIA; Ministerio de Ciencia y Innovation Agencia Estatal de Investigaciones (PGC2018-097027-B-100/10.13039/501100011033, CEX2019-000910-S/10.13039/501100011033, Plan National FIDEUA PID2019-106901GB-I00, FPI, QUANTERA MAQS PCI2019-111828-2, QUANTERA DYNAMITE PCI2022-132919,  Proyectos de I+D+I “Retos Colaboraci\'on” QUSPIN RTC2019-007196-7); MICIIN with funding from European Union NextGenerationEU(PRTR-C17.I1) and by Generalitat de Catalunya;  Fundació Cellex; Fundació Mir-Puig; Generalitat de Catalunya (European social fund FEDER and CERCA program, AGAUR Grant No. 2021 SGR 01452, QuantumCAT / U16-011424, co-funded by ERDF Operational Program of Catalonia 2014-2020); EU Horizon 2020 FET-OPEN OPTOlogic (Grant No 899794); EU Horizon Europe Program (Grant Agreement 101080086 — NeQST), National Science Centre, Poland (Symfonia Grant No. 2016/20/W/ST4/00314); ICFO Internal “QuantumGaudi” project; European Union’s Horizon 2020 research and innovation program under the Marie-Skłodowska-Curie grant agreement No 101029393 (STREDCH) and No 847648  (“La Caixa” Junior Leaders fellowships ID100010434: LCF/BQ/PI19/11690013, LCF/BQ/PI20/11760031,  LCF/BQ/PR20/11770012, LCF/BQ/PR21/11840013). D.G.-C. is supported by the Simons Collaboration on Ultra-Quantum Matter, which is a grant from the Simons Foundation (651440, P.Z.). Views and opinions expressed in this work are, however, those of the author(s) only and do not necessarily reflect those of the European Union, European Climate, Infrastructure and Environment Executive Agency (CINEA), nor any other granting authority. Neither the European Union nor any granting authority can be held responsible for them. The authors thankfully acknowledge the computer resources at MareNostrum and the technical support provided by Barcelona Supercomputing Center (FI-2022-1-0042, FI-2022-3-0039).

\clearpage

\appendix
\section*{Supplementary material}

\section*{Derivation of the effective Hamiltonian}
\label{app:eff_H}

Here we derive Hamiltonians in Eqs. (1) and (2) of the main text, following Refs.~\cite{maschler2005, chanda2021, Ritsch_2013}. We consider a system of ultracold atoms confined to the $(x,y)$-plane by an optical trap, placed within two perpendicular cavities [Fig.1(a) of the main text]. Specifically, the two cavities are single-mode, equal and decoupled, and they are aligned in the $\hat{x}$ and $\hat{y}$ directions, respectively. We consider two internal atomic levels, with transition frequency $\omega_a$, and standing cavity modes with frequency $\omega_c \ll \omega_a$. Finally, we drive the cavity using a transverse laser pump with frequency $\omega_p \approx \omega_c$. In the rotating-wave approximation, the single-particle Hamiltonian reads
\begin{equation}\label{eq:H_JC}
\begin{aligned}
    H^{\rm (sp)} &= \frac{\vec{p}^2}{2m} - \Delta_a \sigma^+\sigma^- + i h_0 \zeta (\sigma^+ - \sigma^-) \\
    & - \Delta_c \sum_\mu a_\mu^\dagger a^{\vphantom{\dagger}}_\mu + i \sum_\mu g(\mu)(\sigma^+a_\mu-\sigma^-a_\mu^\dagger),
\end{aligned}
\end{equation}
with total momentum $\vec{p}=(p_x, p_y)$, and we took $\hbar=1$ and. Here, $\Delta_a = w_p-w_a$ is the atom-pump detuning and $\Delta_c = w_p - w_c$ is the cavity-pump detuning. Moreover, $\sigma^+ (\sigma^-)$ are the raising (lowering) operators for the two-level atoms, and $a_\mu$ ($a_\mu^\dagger$) are the creation (annihilation) operators for the cavity mode $\mu \in \{x,y\}$. Along the $\hat{x}$ and $\hat{y}$ directions, the atom-field coupling is set to $g(x) = g_0 \cos(kx+\phi)$ and $g(y) = g_0 \cos(ky+\phi)$, respectively. On the other hand, the transverse laser field with amplitude $\zeta$ forms a standing wave $h(z) = h_0 \cos(k_p z)$, resulting in $h_0$ in the $z=0$ plane.

We assume low saturation ($\Delta_a \gg h_0, g_0$), which allows us to eliminate the excited atomic state using perturbation theory. Taking
\begin{eqnarray}
    H_0 = \frac{\vec{p}^2}{2m} - \Delta_a \sigma^+\sigma^- - \Delta_c \sum_\mu a_\mu^\dagger a_\mu,
\end{eqnarray}
and
\begin{eqnarray}
    V &=& h_0 \zeta (\sigma^+ - \sigma^-) + i\sum_\mu g(\mu)(\sigma^+a_\mu-\sigma^-a_\mu^\dagger),
\end{eqnarray}
the effective Hamiltonian is given by
\begin{equation}
    H^{\rm (sp)}_{\rm eff} = P_g H_0 P_g + P_g V P_g - P_g V K V P_g,
\end{equation}
with $P_g = \sigma^-\sigma^+$. Explicitly, 
\begin{equation}\label{eq:H_eff}
\begin{aligned}
H^{\rm (sp)}_{\rm eff} &= \frac{\vec{p}^2}{2m} + V_{0}(x,y) + U_0 \sum_\mu \cos^2(k\mu+\phi) a_\mu^\dagger a^{\vphantom{\dagger}}_\mu \\
    & - U_0 \cos(kx+\phi)\cos(ky+\phi)(a_x^\dagger a^{\vphantom{\dagger}}_y + a_y^\dagger a^{\vphantom{\dagger}}_x) \\
    & - \Delta_c \sum_\mu a_\mu^\dagger a^{\vphantom{\dagger}}_\mu + U_p \sum_\mu \cos(k\mu+\phi) (a^{\vphantom{\dagger}}_\mu + a_\mu^\dagger),
\end{aligned}
\end{equation}
where $U_p = g_0 \zeta/\Delta_a$ and $U_0 = g_0^2/\Delta_a$. In the previous expression, we also added an extra square optical lattice [Fig.1(a) of the main text],
\begin{equation}
    V_{0}(x,y) = V_0 [{\rm cos}^2(kx)+{\rm cos}^2(ky)],
\end{equation}
with lattice spacing $\lambda/2$ (where $\lambda=2\pi/k$). Nottice that the lattice depth fluctuates with $U_0 a_\mu^\dagger a^{\vphantom{\dagger}}_\mu$, which will lead to a constant renormalization after we adiabatically eliminate the cavity modes.

We now derive the second-quantized Bose-Hubbard Hamiltonian in Eq. (1) of the main text by combining the single-particle Hamiltonian with two-body atomic contact interactions as follows,
\begin{eqnarray}
    H &=& \int \! d^2\vec{r} \, \psi^\dagger(\vec{r}) H^{\rm (sp)}_{\rm eff} \psi(\vec{r}) 
    \nonumber \\ 
    && + g_s \int \! d^2\vec{r} \, \psi^\dagger(\vec{r}) \psi^\dagger(\vec{r}) \psi(\vec{r})\psi(\vec{r}),
\end{eqnarray}
 We expand $\psi(\vec{r})$ in terms of Wannier functions $\omega_i(\vec{r}-\vec{r}_i)$ centered around the sites $i$ of a square lattice, located at positions $\vec{r}_i = (x_i, y_i)$. Explicitly,  $\psi(\vec{r}) = \sum_i b_i \omega_i(\vec{r}-\vec{r}_i)$, where $b_i$ is the boson creation operator at a $i$ and we restricted ourselves to the lowest vibrational state at each site (tight-binding approximation). By substituting this expression in the second-quantized Hamiltonian and keeping only on-site interactions and nearest-neighbor tunnelings, we obtain the following expression,
 \begin{equation}
 \begin{aligned}
 H =& -t\sum_{\langle i,j \rangle} (b_i^\dagger b^{\vphantom{\dagger}}_j + \text{H.c.}) + \frac{U}{2}\sum_j n_j (n_j -1) \\
    & - \sum_\mu(\Delta_c + \sum_{i,j} V^\mu_{i,j}  b^\dagger_i b^{\vphantom{\dagger}}_j) a_\mu^\dagger a^{\vphantom{\dagger}}_\mu  \\
    &  + \sum_{ij} V_{i,j}^{xy} b_i^\dagger b_j (a_x^\dagger a^{\vphantom{\dagger}}_y + a_y^\dagger a^{\vphantom{\dagger}}_x) \\
    & + \sum_\mu \sum_{i,j} g_{i,j}^\mu (a^{\vphantom{\dagger}}_\mu + a^\dagger_\mu) b_i^\dagger b^{\vphantom{\dagger}}_j,
 \end{aligned}  
 \end{equation}
where the hopping amplitude is given by
\begin{equation}
    t = -\int \! d\vec{r} \, \omega(\vec{r}-\vec{r}_i)\left[\frac{\vec{p}^2}{2m} + V_{0}(x,y)\right]\omega(\vec{r}-\vec{r}_j),
\end{equation}
and the onsite interaction amplitude is
\begin{equation}
    U = g_s \int \! d\vec{r} \, \vert \omega(\vec{r}-\vec{r}_i)\vert^4.
\end{equation}
The amplitudes $g_{i,j}^\mu$ depend explicitly on the period and the phase of the cavity mode, which, as we will show, can be tuned to obtain the desired Hamiltonian. Explicitly,
\begin{eqnarray}
    && g_{i,j}^\mu = U_p \int \! d^2\vec{r} \, \omega(\vec{r}-\vec{r}_i) \cos(k\mu+\phi) \omega(\vec{r}-\vec{r}_i) \\
    && = (-1)^{\mu_i} \frac{\pi}{k}\int d^2 \vec{r} \omega\left(\frac{\pi}{k}\vec{r}\right)\omega\left(\frac{\pi}{k}(\vec{r} + \vec{r}_j - \vec{r}_i)\right) \cos(\pi \mu + \phi). \nonumber
\end{eqnarray}
We take $\phi=\pi/2$, such that lattice sites are located at the nodes of the cavity modes, leading to
\begin{equation}
    \cos(\pi \mu + \frac{\pi}{2}) = -\sin(\pi \mu).
\end{equation}
We can check that due to symmetry arguments, the only surviving terms will be those for which $\vec{r}_j = \vec{r}_i + \hat{\mu}$, which take the form
\begin{equation}
    g_{i,i+\hat{\mu}}^\mu =  (-1)^{\mu_i+1} \frac{\pi}{k}\int \! d^2 \vec{r} \, \omega\left(\frac{\pi}{k}\vec{r} \right)\omega\left(\frac{\pi}{k}(\vec{r} + \hat{\mu})\right) \sin(\pi \mu), \nonumber
\end{equation}
satisfying $g_{i,i+\hat{\mu}}^\mu =  (-1)^{\mu_i+1} g$. We look now at the term 
\begin{equation}
    V_{i,j}^\mu = U_0\int \! d^2 \vec{r} \, \omega(\vec{r}-\vec{r}_i)  \omega(\vec{r}-\vec{r}_j) \sin² (\pi \mu),
\end{equation}    
which, for $i=j$ gives a non-vanishing constant, $V_{i,i}^\mu = V$, and for $\vec{r}_j = \vec{r}_i + \hat{\mu}$ leads to 
\begin{equation}
    V_{i,i+\hat{\mu}}^\mu = U_0 \frac{\pi}{k} \int \! d^2 \vec{r}  \, \omega(\vec{r})  \omega(\vec{r}-\hat{\mu}) \sin^2 (\pi \mu),
\end{equation}
which can be neglected in the limit $g_0 \ll \zeta$. Notice that, since particle number is conserved, the former just contributes to a renormalization of $\Delta_c$. Finally, the terms
\begin{eqnarray}
    V_{i,j}^{xy} &&= U_0\int \! d^2 \, \vec{r}  \omega(\vec{r}-\vec{r}_i)  \omega(\vec{r}-\vec{r}_j) \sin² (\pi x) \sin² (\pi y) \nonumber \\
    && = U_0 (-1)^{x_i+x_j}\frac{\pi}{k} \int \! d^2 \vec{r} \,  \omega\left(\frac{\pi}{k} \vec{r}\right)  \omega\left(\frac{\pi}{k}(\vec{r}+\vec{r}_i-\vec{r}_j )\right) \nonumber \\
    &&\times\sin² (\pi x) \sin² (\pi y)
\end{eqnarray}
only have a non-zero contribution for $\vec{r}_j = \vec{r}_i + \hat{\mu}$, which again can be neglected for $g_0 \ll \zeta$.

The resulting Hamiltonian can be found in Eq. (1) of the main text. We now derive the Hamiltonian in Eq.(2), by adiabatically eliminating the cavity degrees of freedom, for which we assume a large cavity decay rate $k$. In particular, we can write
\begin{equation}
    \dot{a}_\mu = i [k, a_\mu] - ka_\mu,
\end{equation}
which translates to
 \begin{equation}
    \dot{a}_\mu = i [(\Delta_c - V_1 N) a_\mu - ig B - k a_\mu,
\end{equation}
where $N = \sum_i b_i^\dagger b^{\vphantom{\dagger}}_i$ and $B_\mu = \sum_i (-1)^i \left(b_i^\dagger b^{\vphantom{\dagger}}_{i+\hat{\mu}} + \text{H.c.}\right)$. If one assumes a stationary field, i.e. $\dot{a}_\mu = 0$, then
\begin{equation}
    a_\mu = \frac{g B_\mu}{\Delta_c - V_1 N + ik},
\end{equation}
resulting in Eq. (2) of the main text in the limit of large detuning $\Delta_c$.

\section*{Scaling with the bond dimension}
\begin{figure}
\includegraphics[width=1\columnwidth]{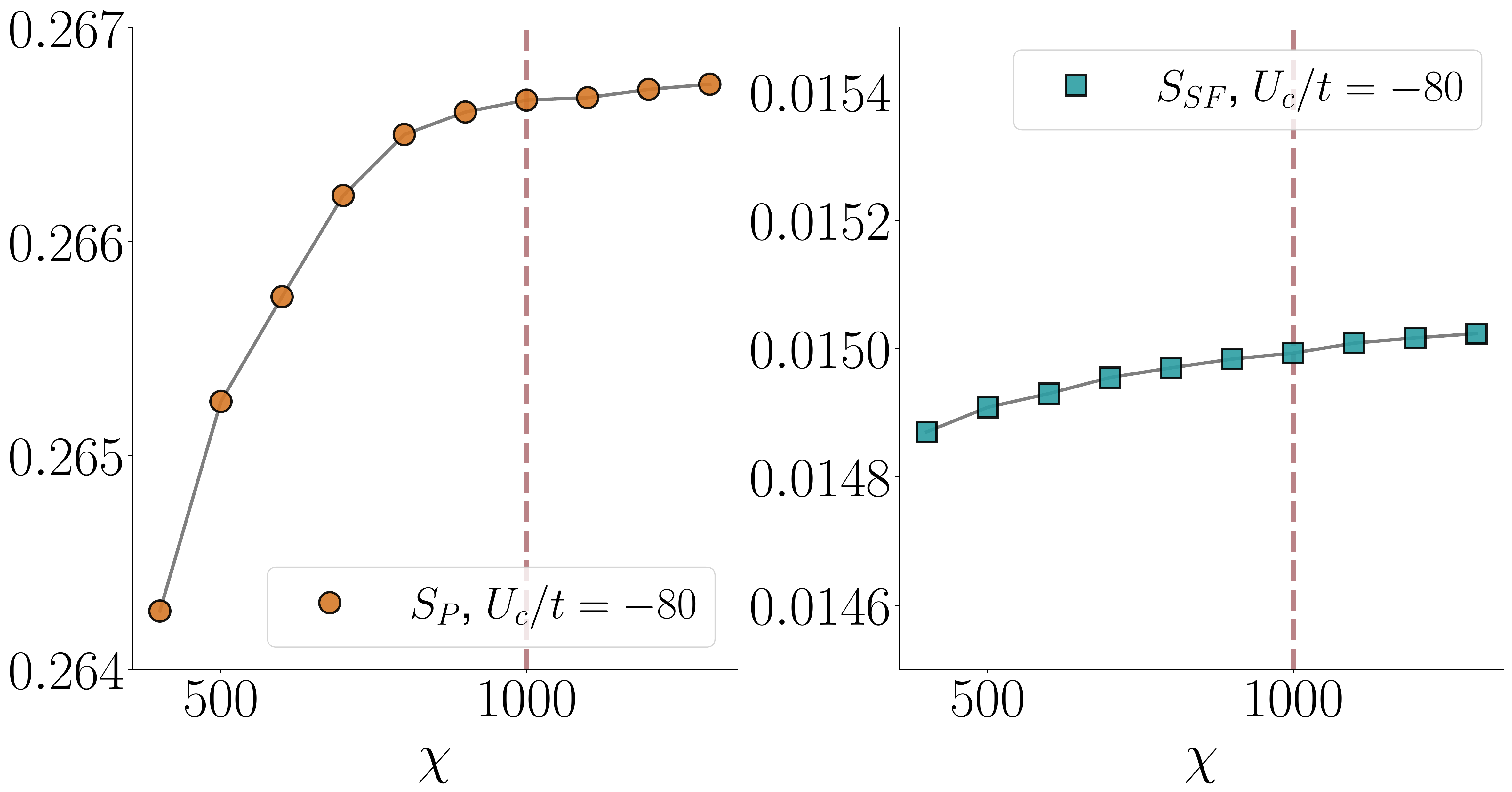}
\caption{\label{fig:figAP} \textbf{Study of the bond dimension}. Scaling of order parameters $S_{\rm SF}$ and $S_{\rm P}$ for a system with size $L^2 = 64$ and coupling $U_c/t = -80$. We see that the two quantities converge for $\chi \approx 1000$. }
\end{figure}
We finally include a study of the convergence with the bond dimension $\chi$ of the superfluid order parameter $S_{\rm SF}$, -defined in Eq.(3) of the main text- and the plaquette order parameter $S_{\rm P}$ -defined in Eq.(4) of the main text.  The scaling of the two quantities is computed for a system of size $L\times L$ with $L = 8$, at $U_c/t=-80$, therefore within the HOTPI phase as shown in Fig.2(a) of the main text. We see that the two order parameters start converging around $\chi \approx 1000$, which is the bond dimension used throughout the paper. Such bond dimension is therefore enough to differentiate between the two regimes and compute the quantities needed to characterize the HOTPI phase.


\begin{thebibliography}{81}%
\makeatletter
\providecommand \@ifxundefined [1]{%
 \@ifx{#1\undefined}
}%
\providecommand \@ifnum [1]{%
 \ifnum #1\expandafter \@firstoftwo
 \else \expandafter \@secondoftwo
 \fi
}%
\providecommand \@ifx [1]{%
 \ifx #1\expandafter \@firstoftwo
 \else \expandafter \@secondoftwo
 \fi
}%
\providecommand \natexlab [1]{#1}%
\providecommand \enquote  [1]{``#1''}%
\providecommand \bibnamefont  [1]{#1}%
\providecommand \bibfnamefont [1]{#1}%
\providecommand \citenamefont [1]{#1}%
\providecommand \href@noop [0]{\@secondoftwo}%
\providecommand \href [0]{\begingroup \@sanitize@url \@href}%
\providecommand \@href[1]{\@@startlink{#1}\@@href}%
\providecommand \@@href[1]{\endgroup#1\@@endlink}%
\providecommand \@sanitize@url [0]{\catcode `\\12\catcode `\$12\catcode
  `\&12\catcode `\#12\catcode `\^12\catcode `\_12\catcode `\%12\relax}%
\providecommand \@@startlink[1]{}%
\providecommand \@@endlink[0]{}%
\providecommand \url  [0]{\begingroup\@sanitize@url \@url }%
\providecommand \@url [1]{\endgroup\@href {#1}{\urlprefix }}%
\providecommand \urlprefix  [0]{URL }%
\providecommand \Eprint [0]{\href }%
\providecommand \doibase [0]{https://doi.org/}%
\providecommand \selectlanguage [0]{\@gobble}%
\providecommand \bibinfo  [0]{\@secondoftwo}%
\providecommand \bibfield  [0]{\@secondoftwo}%
\providecommand \translation [1]{[#1]}%
\providecommand \BibitemOpen [0]{}%
\providecommand \bibitemStop [0]{}%
\providecommand \bibitemNoStop [0]{.\EOS\space}%
\providecommand \EOS [0]{\spacefactor3000\relax}%
\providecommand \BibitemShut  [1]{\csname bibitem#1\endcsname}%
\let\auto@bib@innerbib\@empty
\bibitem [{\citenamefont {Moore}(2010)}]{Moore_2010}%
  \BibitemOpen
  \bibfield  {author} {\bibinfo {author} {\bibfnamefont {J.~E.}\ \bibnamefont
  {Moore}},\ }\href {https://doi.org/10.1038/nature08916} {\bibfield  {journal}
  {\bibinfo  {journal} {Nature}\ }\textbf {\bibinfo {volume} {464}},\ \bibinfo
  {pages} {194} (\bibinfo {year} {2010})}\BibitemShut {NoStop}%
\bibitem [{\citenamefont {Landau}(1937)}]{Landau2937}%
  \BibitemOpen
  \bibfield  {author} {\bibinfo {author} {\bibfnamefont {L.~D.}\ \bibnamefont
  {Landau}},\ }\href {https://doi.org/10.1016/B978-0-08-010586-4.50034-1}
  {\bibfield  {journal} {\bibinfo  {journal} {Zh. Eksp. Teor. Fiz.}\ }\textbf
  {\bibinfo {volume} {7}},\ \bibinfo {pages} {19} (\bibinfo {year}
  {1937})}\BibitemShut {NoStop}%
\bibitem [{\citenamefont {Hasan}\ and\ \citenamefont
  {Kane}(2010)}]{Hasan_2010}%
  \BibitemOpen
  \bibfield  {author} {\bibinfo {author} {\bibfnamefont {M.~Z.}\ \bibnamefont
  {Hasan}}\ and\ \bibinfo {author} {\bibfnamefont {C.~L.}\ \bibnamefont
  {Kane}},\ }\href {https://doi.org/10.1103/RevModPhys.82.3045} {\bibfield
  {journal} {\bibinfo  {journal} {Rev. Mod. Phys.}\ }\textbf {\bibinfo {volume}
  {82}},\ \bibinfo {pages} {3045} (\bibinfo {year} {2010})}\BibitemShut
  {NoStop}%
\bibitem [{\citenamefont {Qi}\ and\ \citenamefont {Zhang}(2011)}]{Qi_2011}%
  \BibitemOpen
  \bibfield  {author} {\bibinfo {author} {\bibfnamefont {X.-L.}\ \bibnamefont
  {Qi}}\ and\ \bibinfo {author} {\bibfnamefont {S.-C.}\ \bibnamefont {Zhang}},\
  }\href {https://doi.org/10.1103/RevModPhys.83.1057} {\bibfield  {journal}
  {\bibinfo  {journal} {Rev. Mod. Phys.}\ }\textbf {\bibinfo {volume} {83}},\
  \bibinfo {pages} {1057} (\bibinfo {year} {2011})}\BibitemShut {NoStop}%
\bibitem [{\citenamefont {MacDonald}(1990)}]{MacDonald_1990}%
  \BibitemOpen
  \bibfield  {author} {\bibinfo {author} {\bibfnamefont {A.~H.}\ \bibnamefont
  {MacDonald}},\ }\href {https://doi.org/10.1103/PhysRevLett.64.220} {\bibfield
   {journal} {\bibinfo  {journal} {Phys. Rev. Lett.}\ }\textbf {\bibinfo
  {volume} {64}},\ \bibinfo {pages} {220} (\bibinfo {year} {1990})}\BibitemShut
  {NoStop}%
\bibitem [{\citenamefont {Hatsugai}(1993)}]{Hatsugai_1993}%
  \BibitemOpen
  \bibfield  {author} {\bibinfo {author} {\bibfnamefont {Y.}~\bibnamefont
  {Hatsugai}},\ }\href {https://doi.org/10.1103/PhysRevLett.71.3697} {\bibfield
   {journal} {\bibinfo  {journal} {Phys. Rev. Lett.}\ }\textbf {\bibinfo
  {volume} {71}},\ \bibinfo {pages} {3697} (\bibinfo {year}
  {1993})}\BibitemShut {NoStop}%
\bibitem [{\citenamefont {Ryu}\ and\ \citenamefont
  {Hatsugai}(2002)}]{Ryu_2002}%
  \BibitemOpen
  \bibfield  {author} {\bibinfo {author} {\bibfnamefont {S.}~\bibnamefont
  {Ryu}}\ and\ \bibinfo {author} {\bibfnamefont {Y.}~\bibnamefont {Hatsugai}},\
  }\href {https://doi.org/10.1103/PhysRevLett.89.077002} {\bibfield  {journal}
  {\bibinfo  {journal} {Phys. Rev. Lett.}\ }\textbf {\bibinfo {volume} {89}},\
  \bibinfo {pages} {077002} (\bibinfo {year} {2002})}\BibitemShut {NoStop}%
\bibitem [{\citenamefont {Klitzing}\ \emph {et~al.}(1980)\citenamefont
  {Klitzing}, \citenamefont {Dorda},\ and\ \citenamefont {Pepper}}]{von}%
  \BibitemOpen
  \bibfield  {author} {\bibinfo {author} {\bibfnamefont {K.~v.}\ \bibnamefont
  {Klitzing}}, \bibinfo {author} {\bibfnamefont {G.}~\bibnamefont {Dorda}},\
  and\ \bibinfo {author} {\bibfnamefont {M.}~\bibnamefont {Pepper}},\ }\href
  {https://doi.org/10.1103/PhysRevLett.45.494} {\bibfield  {journal} {\bibinfo
  {journal} {Phys. Rev. Lett.}\ }\textbf {\bibinfo {volume} {45}},\ \bibinfo
  {pages} {494} (\bibinfo {year} {1980})}\BibitemShut {NoStop}%
\bibitem [{\citenamefont {Laughlin}(1981)}]{laughlin}%
  \BibitemOpen
  \bibfield  {author} {\bibinfo {author} {\bibfnamefont {R.~B.}\ \bibnamefont
  {Laughlin}},\ }\href {https://doi.org/10.1103/PhysRevB.23.5632} {\bibfield
  {journal} {\bibinfo  {journal} {Phys. Rev. B}\ }\textbf {\bibinfo {volume}
  {23}},\ \bibinfo {pages} {5632} (\bibinfo {year} {1981})}\BibitemShut
  {NoStop}%
\bibitem [{\citenamefont {Thouless}\ \emph {et~al.}(1982)\citenamefont
  {Thouless}, \citenamefont {Kohmoto}, \citenamefont {Nightingale},\ and\
  \citenamefont {den Nijs}}]{thouless}%
  \BibitemOpen
  \bibfield  {author} {\bibinfo {author} {\bibfnamefont {D.~J.}\ \bibnamefont
  {Thouless}}, \bibinfo {author} {\bibfnamefont {M.}~\bibnamefont {Kohmoto}},
  \bibinfo {author} {\bibfnamefont {M.~P.}\ \bibnamefont {Nightingale}},\ and\
  \bibinfo {author} {\bibfnamefont {M.}~\bibnamefont {den Nijs}},\ }\href
  {https://doi.org/10.1103/PhysRevLett.49.405} {\bibfield  {journal} {\bibinfo
  {journal} {Phys. Rev. Lett.}\ }\textbf {\bibinfo {volume} {49}},\ \bibinfo
  {pages} {405} (\bibinfo {year} {1982})}\BibitemShut {NoStop}%
\bibitem [{\citenamefont {Read}\ and\ \citenamefont {Moore}(1992)}]{moore}%
  \BibitemOpen
  \bibfield  {author} {\bibinfo {author} {\bibfnamefont {N.}~\bibnamefont
  {Read}}\ and\ \bibinfo {author} {\bibfnamefont {G.}~\bibnamefont {Moore}},\
  }\href {https://doi.org/10.1143/ptps.107.157} {\bibfield  {journal} {\bibinfo
   {journal} {Progress of Theoretical Physics Supplement}\ }\textbf {\bibinfo
  {volume} {107}},\ \bibinfo {pages} {157–166} (\bibinfo {year}
  {1992})}\BibitemShut {NoStop}%
\bibitem [{\citenamefont {Tsui}\ \emph {et~al.}(1982)\citenamefont {Tsui},
  \citenamefont {Stormer},\ and\ \citenamefont {Gossard}}]{tsui}%
  \BibitemOpen
  \bibfield  {author} {\bibinfo {author} {\bibfnamefont {D.~C.}\ \bibnamefont
  {Tsui}}, \bibinfo {author} {\bibfnamefont {H.~L.}\ \bibnamefont {Stormer}},\
  and\ \bibinfo {author} {\bibfnamefont {A.~C.}\ \bibnamefont {Gossard}},\
  }\href {https://doi.org/10.1103/PhysRevLett.48.1559} {\bibfield  {journal}
  {\bibinfo  {journal} {Phys. Rev. Lett.}\ }\textbf {\bibinfo {volume} {48}},\
  \bibinfo {pages} {1559} (\bibinfo {year} {1982})}\BibitemShut {NoStop}%
\bibitem [{\citenamefont {Laughlin}(1983)}]{laughlin1}%
  \BibitemOpen
  \bibfield  {author} {\bibinfo {author} {\bibfnamefont {R.~B.}\ \bibnamefont
  {Laughlin}},\ }\href {https://doi.org/10.1103/PhysRevLett.50.1395} {\bibfield
   {journal} {\bibinfo  {journal} {Phys. Rev. Lett.}\ }\textbf {\bibinfo
  {volume} {50}},\ \bibinfo {pages} {1395} (\bibinfo {year}
  {1983})}\BibitemShut {NoStop}%
\bibitem [{\citenamefont {Su}\ and\ \citenamefont {Schrieffer}(1981)}]{su}%
  \BibitemOpen
  \bibfield  {author} {\bibinfo {author} {\bibfnamefont {W.~P.}\ \bibnamefont
  {Su}}\ and\ \bibinfo {author} {\bibfnamefont {J.~R.}\ \bibnamefont
  {Schrieffer}},\ }\href {https://doi.org/10.1103/PhysRevLett.46.738}
  {\bibfield  {journal} {\bibinfo  {journal} {Phys. Rev. Lett.}\ }\textbf
  {\bibinfo {volume} {46}},\ \bibinfo {pages} {738} (\bibinfo {year}
  {1981})}\BibitemShut {NoStop}%
\bibitem [{\citenamefont {Chiu}\ \emph {et~al.}(2016)\citenamefont {Chiu},
  \citenamefont {Teo}, \citenamefont {Schnyder},\ and\ \citenamefont
  {Ryu}}]{chiu2016}%
  \BibitemOpen
  \bibfield  {author} {\bibinfo {author} {\bibfnamefont {C.-K.}\ \bibnamefont
  {Chiu}}, \bibinfo {author} {\bibfnamefont {J.~C.~Y.}\ \bibnamefont {Teo}},
  \bibinfo {author} {\bibfnamefont {A.~P.}\ \bibnamefont {Schnyder}},\ and\
  \bibinfo {author} {\bibfnamefont {S.}~\bibnamefont {Ryu}},\ }\href
  {https://doi.org/10.1103/RevModPhys.88.035005} {\bibfield  {journal}
  {\bibinfo  {journal} {Rev. Mod. Phys.}\ }\textbf {\bibinfo {volume} {88}},\
  \bibinfo {pages} {035005} (\bibinfo {year} {2016})}\BibitemShut {NoStop}%
\bibitem [{\citenamefont {Benalcazar}\ \emph
  {et~al.}(2017{\natexlab{a}})\citenamefont {Benalcazar}, \citenamefont
  {Bernevig},\ and\ \citenamefont {Hughes}}]{benalcazar2017}%
  \BibitemOpen
  \bibfield  {author} {\bibinfo {author} {\bibfnamefont {W.~A.}\ \bibnamefont
  {Benalcazar}}, \bibinfo {author} {\bibfnamefont {B.~A.}\ \bibnamefont
  {Bernevig}},\ and\ \bibinfo {author} {\bibfnamefont {T.~L.}\ \bibnamefont
  {Hughes}},\ }\href {https://doi.org/10.1126/science.aah6442} {\bibfield
  {journal} {\bibinfo  {journal} {Science}\ }\textbf {\bibinfo {volume}
  {357}},\ \bibinfo {pages} {61} (\bibinfo {year}
  {2017}{\natexlab{a}})}\BibitemShut {NoStop}%
\bibitem [{\citenamefont {Benalcazar}\ \emph
  {et~al.}(2017{\natexlab{b}})\citenamefont {Benalcazar}, \citenamefont
  {Bernevig},\ and\ \citenamefont {Hughes}}]{Benalcaza_2017_2}%
  \BibitemOpen
  \bibfield  {author} {\bibinfo {author} {\bibfnamefont {W.~A.}\ \bibnamefont
  {Benalcazar}}, \bibinfo {author} {\bibfnamefont {B.~A.}\ \bibnamefont
  {Bernevig}},\ and\ \bibinfo {author} {\bibfnamefont {T.~L.}\ \bibnamefont
  {Hughes}},\ }\href {https://doi.org/10.1103/PhysRevB.96.245115} {\bibfield
  {journal} {\bibinfo  {journal} {Phys. Rev. B}\ }\textbf {\bibinfo {volume}
  {96}},\ \bibinfo {pages} {245115} (\bibinfo {year}
  {2017}{\natexlab{b}})}\BibitemShut {NoStop}%
\bibitem [{\citenamefont {Schindler}\ \emph {et~al.}(2018)\citenamefont
  {Schindler}, \citenamefont {Cook}, \citenamefont {Vergniory}, \citenamefont
  {Wang}, \citenamefont {Parkin}, \citenamefont {Bernevig},\ and\ \citenamefont
  {Neupert}}]{schindler2018}%
  \BibitemOpen
  \bibfield  {author} {\bibinfo {author} {\bibfnamefont {F.}~\bibnamefont
  {Schindler}}, \bibinfo {author} {\bibfnamefont {A.~M.}\ \bibnamefont {Cook}},
  \bibinfo {author} {\bibfnamefont {M.~G.}\ \bibnamefont {Vergniory}}, \bibinfo
  {author} {\bibfnamefont {Z.}~\bibnamefont {Wang}}, \bibinfo {author}
  {\bibfnamefont {S.~S.~P.}\ \bibnamefont {Parkin}}, \bibinfo {author}
  {\bibfnamefont {B.~A.}\ \bibnamefont {Bernevig}},\ and\ \bibinfo {author}
  {\bibfnamefont {T.}~\bibnamefont {Neupert}},\ }\bibfield  {journal} {\bibinfo
   {journal} {Science Advances}\ }\textbf {\bibinfo {volume} {4}},\ \href
  {https://doi.org/10.1126/sciadv.aat0346} {10.1126/sciadv.aat0346} (\bibinfo
  {year} {2018})\BibitemShut {NoStop}%
\bibitem [{\citenamefont {Khalaf}(2018)}]{Khalaf_2018}%
  \BibitemOpen
  \bibfield  {author} {\bibinfo {author} {\bibfnamefont {E.}~\bibnamefont
  {Khalaf}},\ }\href {https://doi.org/10.1103/PhysRevB.97.205136} {\bibfield
  {journal} {\bibinfo  {journal} {Phys. Rev. B}\ }\textbf {\bibinfo {volume}
  {97}},\ \bibinfo {pages} {205136} (\bibinfo {year} {2018})}\BibitemShut
  {NoStop}%
\bibitem [{\citenamefont {You}\ \emph {et~al.}(2018)\citenamefont {You},
  \citenamefont {Devakul}, \citenamefont {Burnell},\ and\ \citenamefont
  {Neupert}}]{You_2018}%
  \BibitemOpen
  \bibfield  {author} {\bibinfo {author} {\bibfnamefont {Y.}~\bibnamefont
  {You}}, \bibinfo {author} {\bibfnamefont {T.}~\bibnamefont {Devakul}},
  \bibinfo {author} {\bibfnamefont {F.~J.}\ \bibnamefont {Burnell}},\ and\
  \bibinfo {author} {\bibfnamefont {T.}~\bibnamefont {Neupert}},\ }\href
  {https://doi.org/10.1103/PhysRevB.98.235102} {\bibfield  {journal} {\bibinfo
  {journal} {Phys. Rev. B}\ }\textbf {\bibinfo {volume} {98}},\ \bibinfo
  {pages} {235102} (\bibinfo {year} {2018})}\BibitemShut {NoStop}%
\bibitem [{\citenamefont {Dubinkin}\ and\ \citenamefont
  {Hughes}(2019)}]{Dubinkin_2019}%
  \BibitemOpen
  \bibfield  {author} {\bibinfo {author} {\bibfnamefont {O.}~\bibnamefont
  {Dubinkin}}\ and\ \bibinfo {author} {\bibfnamefont {T.~L.}\ \bibnamefont
  {Hughes}},\ }\href {https://doi.org/10.1103/PhysRevB.99.235132} {\bibfield
  {journal} {\bibinfo  {journal} {Phys. Rev. B}\ }\textbf {\bibinfo {volume}
  {99}},\ \bibinfo {pages} {235132} (\bibinfo {year} {2019})}\BibitemShut
  {NoStop}%
\bibitem [{\citenamefont {Kudo}\ \emph {et~al.}(2019)\citenamefont {Kudo},
  \citenamefont {Yoshida},\ and\ \citenamefont {Hatsugai}}]{Kudo_2019}%
  \BibitemOpen
  \bibfield  {author} {\bibinfo {author} {\bibfnamefont {K.}~\bibnamefont
  {Kudo}}, \bibinfo {author} {\bibfnamefont {T.}~\bibnamefont {Yoshida}},\ and\
  \bibinfo {author} {\bibfnamefont {Y.}~\bibnamefont {Hatsugai}},\ }\href
  {https://doi.org/10.1103/PhysRevLett.123.196402} {\bibfield  {journal}
  {\bibinfo  {journal} {Phys. Rev. Lett.}\ }\textbf {\bibinfo {volume} {123}},\
  \bibinfo {pages} {196402} (\bibinfo {year} {2019})}\BibitemShut {NoStop}%
\bibitem [{\citenamefont {Laubscher}\ \emph {et~al.}(2019)\citenamefont
  {Laubscher}, \citenamefont {Loss},\ and\ \citenamefont
  {Klinovaja}}]{Laubscher_2019}%
  \BibitemOpen
  \bibfield  {author} {\bibinfo {author} {\bibfnamefont {K.}~\bibnamefont
  {Laubscher}}, \bibinfo {author} {\bibfnamefont {D.}~\bibnamefont {Loss}},\
  and\ \bibinfo {author} {\bibfnamefont {J.}~\bibnamefont {Klinovaja}},\ }\href
  {https://doi.org/10.1103/PhysRevResearch.1.032017} {\bibfield  {journal}
  {\bibinfo  {journal} {Phys. Rev. Research}\ }\textbf {\bibinfo {volume}
  {1}},\ \bibinfo {pages} {032017} (\bibinfo {year} {2019})}\BibitemShut
  {NoStop}%
\bibitem [{\citenamefont {Laubscher}\ \emph {et~al.}(2020)\citenamefont
  {Laubscher}, \citenamefont {Loss},\ and\ \citenamefont
  {Klinovaja}}]{Laubscher_2020}%
  \BibitemOpen
  \bibfield  {author} {\bibinfo {author} {\bibfnamefont {K.}~\bibnamefont
  {Laubscher}}, \bibinfo {author} {\bibfnamefont {D.}~\bibnamefont {Loss}},\
  and\ \bibinfo {author} {\bibfnamefont {J.}~\bibnamefont {Klinovaja}},\ }\href
  {https://doi.org/10.1103/PhysRevResearch.2.013330} {\bibfield  {journal}
  {\bibinfo  {journal} {Phys. Rev. Research}\ }\textbf {\bibinfo {volume}
  {2}},\ \bibinfo {pages} {013330} (\bibinfo {year} {2020})}\BibitemShut
  {NoStop}%
\bibitem [{\citenamefont {Sil}\ and\ \citenamefont {Ghosh}(2020)}]{Sil_2020}%
  \BibitemOpen
  \bibfield  {author} {\bibinfo {author} {\bibfnamefont {A.}~\bibnamefont
  {Sil}}\ and\ \bibinfo {author} {\bibfnamefont {A.~K.}\ \bibnamefont
  {Ghosh}},\ }\href {https://doi.org/10.1088/1361-648x/ab6f8b} {\bibfield
  {journal} {\bibinfo  {journal} {Journal of Physics: Condensed Matter}\
  }\textbf {\bibinfo {volume} {32}},\ \bibinfo {pages} {205601} (\bibinfo
  {year} {2020})}\BibitemShut {NoStop}%
\bibitem [{\citenamefont {Rasmussen}\ and\ \citenamefont
  {Lu}(2020)}]{Rasmussen_2020}%
  \BibitemOpen
  \bibfield  {author} {\bibinfo {author} {\bibfnamefont {A.}~\bibnamefont
  {Rasmussen}}\ and\ \bibinfo {author} {\bibfnamefont {Y.-M.}\ \bibnamefont
  {Lu}},\ }\href {https://doi.org/10.1103/PhysRevB.101.085137} {\bibfield
  {journal} {\bibinfo  {journal} {Phys. Rev. B}\ }\textbf {\bibinfo {volume}
  {101}},\ \bibinfo {pages} {085137} (\bibinfo {year} {2020})}\BibitemShut
  {NoStop}%
\bibitem [{\citenamefont {Bibo}\ \emph {et~al.}(2020)\citenamefont {Bibo},
  \citenamefont {Lovas}, \citenamefont {You}, \citenamefont {Grusdt},\ and\
  \citenamefont {Pollmann}}]{bibo2020}%
  \BibitemOpen
  \bibfield  {author} {\bibinfo {author} {\bibfnamefont {J.}~\bibnamefont
  {Bibo}}, \bibinfo {author} {\bibfnamefont {I.}~\bibnamefont {Lovas}},
  \bibinfo {author} {\bibfnamefont {Y.}~\bibnamefont {You}}, \bibinfo {author}
  {\bibfnamefont {F.}~\bibnamefont {Grusdt}},\ and\ \bibinfo {author}
  {\bibfnamefont {F.}~\bibnamefont {Pollmann}},\ }\href
  {https://doi.org/10.1103/PhysRevB.102.041126} {\bibfield  {journal} {\bibinfo
   {journal} {Phys. Rev. B}\ }\textbf {\bibinfo {volume} {102}},\ \bibinfo
  {pages} {041126} (\bibinfo {year} {2020})}\BibitemShut {NoStop}%
\bibitem [{\citenamefont {Peng}\ \emph {et~al.}(2021)\citenamefont {Peng},
  \citenamefont {Zhang},\ and\ \citenamefont {Lu}}]{Peng_2021}%
  \BibitemOpen
  \bibfield  {author} {\bibinfo {author} {\bibfnamefont {C.}~\bibnamefont
  {Peng}}, \bibinfo {author} {\bibfnamefont {L.}~\bibnamefont {Zhang}},\ and\
  \bibinfo {author} {\bibfnamefont {Z.-Y.}\ \bibnamefont {Lu}},\ }\href
  {https://doi.org/10.1103/PhysRevB.104.075112} {\bibfield  {journal} {\bibinfo
   {journal} {Phys. Rev. B}\ }\textbf {\bibinfo {volume} {104}},\ \bibinfo
  {pages} {075112} (\bibinfo {year} {2021})}\BibitemShut {NoStop}%
\bibitem [{\citenamefont {Guo}\ \emph {et~al.}(2021)\citenamefont {Guo},
  \citenamefont {Sun}, \citenamefont {Zhu}, \citenamefont {Li}, \citenamefont
  {Guo},\ and\ \citenamefont {Feng}}]{Guo_2022}%
  \BibitemOpen
  \bibfield  {author} {\bibinfo {author} {\bibfnamefont {J.}~\bibnamefont
  {Guo}}, \bibinfo {author} {\bibfnamefont {J.}~\bibnamefont {Sun}}, \bibinfo
  {author} {\bibfnamefont {X.}~\bibnamefont {Zhu}}, \bibinfo {author}
  {\bibfnamefont {C.-A.}\ \bibnamefont {Li}}, \bibinfo {author} {\bibfnamefont
  {H.}~\bibnamefont {Guo}},\ and\ \bibinfo {author} {\bibfnamefont
  {S.}~\bibnamefont {Feng}},\ }\href {https://doi.org/10.1088/1361-648X/ac30b4}
  {\bibfield  {journal} {\bibinfo  {journal} {Journal of Physics: Condensed
  Matter}\ }\textbf {\bibinfo {volume} {34}},\ \bibinfo {pages} {035603}
  (\bibinfo {year} {2021})}\BibitemShut {NoStop}%
\bibitem [{\citenamefont {Hackenbroich}\ \emph {et~al.}(2021)\citenamefont
  {Hackenbroich}, \citenamefont {Hudomal}, \citenamefont {Schuch},
  \citenamefont {Bernevig},\ and\ \citenamefont
  {Regnault}}]{Hackenbroich_2021}%
  \BibitemOpen
  \bibfield  {author} {\bibinfo {author} {\bibfnamefont {A.}~\bibnamefont
  {Hackenbroich}}, \bibinfo {author} {\bibfnamefont {A.}~\bibnamefont
  {Hudomal}}, \bibinfo {author} {\bibfnamefont {N.}~\bibnamefont {Schuch}},
  \bibinfo {author} {\bibfnamefont {B.~A.}\ \bibnamefont {Bernevig}},\ and\
  \bibinfo {author} {\bibfnamefont {N.}~\bibnamefont {Regnault}},\ }\href
  {https://doi.org/10.1103/PhysRevB.103.L161110} {\bibfield  {journal}
  {\bibinfo  {journal} {Phys. Rev. B}\ }\textbf {\bibinfo {volume} {103}},\
  \bibinfo {pages} {L161110} (\bibinfo {year} {2021})}\BibitemShut {NoStop}%
\bibitem [{\citenamefont {Otsuka}\ \emph {et~al.}(2021)\citenamefont {Otsuka},
  \citenamefont {Yoshida}, \citenamefont {Kudo}, \citenamefont {Yunoki},\ and\
  \citenamefont {Hatsugai}}]{Otsuka_2021}%
  \BibitemOpen
  \bibfield  {author} {\bibinfo {author} {\bibfnamefont {Y.}~\bibnamefont
  {Otsuka}}, \bibinfo {author} {\bibfnamefont {T.}~\bibnamefont {Yoshida}},
  \bibinfo {author} {\bibfnamefont {K.}~\bibnamefont {Kudo}}, \bibinfo {author}
  {\bibfnamefont {S.}~\bibnamefont {Yunoki}},\ and\ \bibinfo {author}
  {\bibfnamefont {Y.}~\bibnamefont {Hatsugai}},\ }\href
  {https://doi.org/10.1038/s41598-021-99213-z} {\bibfield  {journal} {\bibinfo
  {journal} {Scientific Reports}\ }\textbf {\bibinfo {volume} {11}},\ \bibinfo
  {pages} {20270} (\bibinfo {year} {2021})}\BibitemShut {NoStop}%
\bibitem [{\citenamefont {Gonz\'alez-Cuadra}(2022)}]{gonzalez2022}%
  \BibitemOpen
  \bibfield  {author} {\bibinfo {author} {\bibfnamefont {D.}~\bibnamefont
  {Gonz\'alez-Cuadra}},\ }\href {https://doi.org/10.1103/PhysRevB.105.L020403}
  {\bibfield  {journal} {\bibinfo  {journal} {Phys. Rev. B}\ }\textbf {\bibinfo
  {volume} {105}},\ \bibinfo {pages} {L020403} (\bibinfo {year}
  {2022})}\BibitemShut {NoStop}%
\bibitem [{\citenamefont {Li}\ \emph {et~al.}(2022)\citenamefont {Li},
  \citenamefont {Wu}, \citenamefont {Luo}, \citenamefont {Huang},\ and\
  \citenamefont {Chang}}]{Li_2022}%
  \BibitemOpen
  \bibfield  {author} {\bibinfo {author} {\bibfnamefont {Y.-M.}\ \bibnamefont
  {Li}}, \bibinfo {author} {\bibfnamefont {Y.-J.}\ \bibnamefont {Wu}}, \bibinfo
  {author} {\bibfnamefont {X.-W.}\ \bibnamefont {Luo}}, \bibinfo {author}
  {\bibfnamefont {Y.}~\bibnamefont {Huang}},\ and\ \bibinfo {author}
  {\bibfnamefont {K.}~\bibnamefont {Chang}},\ }\href
  {https://doi.org/10.1103/PhysRevB.106.054403} {\bibfield  {journal} {\bibinfo
   {journal} {Phys. Rev. B}\ }\textbf {\bibinfo {volume} {106}},\ \bibinfo
  {pages} {054403} (\bibinfo {year} {2022})}\BibitemShut {NoStop}%
\bibitem [{\citenamefont {Montorsi}\ \emph {et~al.}(2022)\citenamefont
  {Montorsi}, \citenamefont {Bhattacharya}, \citenamefont {Gonz\'alez-Cuadra},
  \citenamefont {Lewenstein}, \citenamefont {Palumbo},\ and\ \citenamefont
  {Barbiero}}]{Montorsi_2022}%
  \BibitemOpen
  \bibfield  {author} {\bibinfo {author} {\bibfnamefont {A.}~\bibnamefont
  {Montorsi}}, \bibinfo {author} {\bibfnamefont {U.}~\bibnamefont
  {Bhattacharya}}, \bibinfo {author} {\bibfnamefont {D.}~\bibnamefont
  {Gonz\'alez-Cuadra}}, \bibinfo {author} {\bibfnamefont {M.}~\bibnamefont
  {Lewenstein}}, \bibinfo {author} {\bibfnamefont {G.}~\bibnamefont
  {Palumbo}},\ and\ \bibinfo {author} {\bibfnamefont {L.}~\bibnamefont
  {Barbiero}},\ }\href {https://doi.org/10.1103/PhysRevB.106.L241115}
  {\bibfield  {journal} {\bibinfo  {journal} {Phys. Rev. B}\ }\textbf {\bibinfo
  {volume} {106}},\ \bibinfo {pages} {L241115} (\bibinfo {year}
  {2022})}\BibitemShut {NoStop}%
\bibitem [{\citenamefont {Wienand}\ \emph {et~al.}(2022)\citenamefont
  {Wienand}, \citenamefont {Horn}, \citenamefont {Aidelsburger}, \citenamefont
  {Bibo},\ and\ \citenamefont {Grusdt}}]{Wienand_2022}%
  \BibitemOpen
  \bibfield  {author} {\bibinfo {author} {\bibfnamefont {J.~F.}\ \bibnamefont
  {Wienand}}, \bibinfo {author} {\bibfnamefont {F.}~\bibnamefont {Horn}},
  \bibinfo {author} {\bibfnamefont {M.}~\bibnamefont {Aidelsburger}}, \bibinfo
  {author} {\bibfnamefont {J.}~\bibnamefont {Bibo}},\ and\ \bibinfo {author}
  {\bibfnamefont {F.}~\bibnamefont {Grusdt}},\ }\href
  {https://doi.org/10.1103/PhysRevLett.128.246602} {\bibfield  {journal}
  {\bibinfo  {journal} {Phys. Rev. Lett.}\ }\textbf {\bibinfo {volume} {128}},\
  \bibinfo {pages} {246602} (\bibinfo {year} {2022})}\BibitemShut {NoStop}%
\bibitem [{\citenamefont {Aksenov}\ \emph {et~al.}(2023)\citenamefont
  {Aksenov}, \citenamefont {Fedoseev}, \citenamefont {Shustin},\ and\
  \citenamefont {Zlotnikov}}]{Aksenov_2023}%
  \BibitemOpen
  \bibfield  {author} {\bibinfo {author} {\bibfnamefont {S.~V.}\ \bibnamefont
  {Aksenov}}, \bibinfo {author} {\bibfnamefont {A.~D.}\ \bibnamefont
  {Fedoseev}}, \bibinfo {author} {\bibfnamefont {M.~S.}\ \bibnamefont
  {Shustin}},\ and\ \bibinfo {author} {\bibfnamefont {A.~O.}\ \bibnamefont
  {Zlotnikov}},\ }\href {https://doi.org/10.1103/PhysRevB.107.125401}
  {\bibfield  {journal} {\bibinfo  {journal} {Phys. Rev. B}\ }\textbf {\bibinfo
  {volume} {107}},\ \bibinfo {pages} {125401} (\bibinfo {year}
  {2023})}\BibitemShut {NoStop}%
\bibitem [{\citenamefont {Su}\ \emph {et~al.}(1979)\citenamefont {Su},
  \citenamefont {Schrieffer},\ and\ \citenamefont {Heeger}}]{ssh}%
  \BibitemOpen
  \bibfield  {author} {\bibinfo {author} {\bibfnamefont {W.~P.}\ \bibnamefont
  {Su}}, \bibinfo {author} {\bibfnamefont {J.~R.}\ \bibnamefont {Schrieffer}},\
  and\ \bibinfo {author} {\bibfnamefont {A.~J.}\ \bibnamefont {Heeger}},\
  }\href {https://doi.org/10.1103/PhysRevLett.42.1698} {\bibfield  {journal}
  {\bibinfo  {journal} {Phys. Rev. Lett.}\ }\textbf {\bibinfo {volume} {42}},\
  \bibinfo {pages} {1698} (\bibinfo {year} {1979})}\BibitemShut {NoStop}%
\bibitem [{\citenamefont {Peierls}(1955)}]{Peierls_1955}%
  \BibitemOpen
  \bibfield  {author} {\bibinfo {author} {\bibfnamefont {R.~E.}\ \bibnamefont
  {Peierls}},\ }\href@noop {} {\emph {\bibinfo {title} {Quantum theory of
  solids}}},\ \bibinfo {number} {23}\ (\bibinfo  {publisher} {Oxford University
  Press},\ \bibinfo {year} {1955})\BibitemShut {NoStop}%
\bibitem [{\citenamefont {Asb{\'o}th}\ \emph {et~al.}(2016)\citenamefont
  {Asb{\'o}th}, \citenamefont {Oroszl{\'a}ny},\ and\ \citenamefont
  {P{\'a}lyi}}]{Asboth_2016}%
  \BibitemOpen
  \bibfield  {author} {\bibinfo {author} {\bibfnamefont {J.~K.}\ \bibnamefont
  {Asb{\'o}th}}, \bibinfo {author} {\bibfnamefont {L.}~\bibnamefont
  {Oroszl{\'a}ny}},\ and\ \bibinfo {author} {\bibfnamefont {A.}~\bibnamefont
  {P{\'a}lyi}},\ }\href@noop {} {\bibfield  {journal} {\bibinfo  {journal}
  {Lecture Notes in Physics}\ }\textbf {\bibinfo {volume} {919}} (\bibinfo
  {year} {2016})}\BibitemShut {NoStop}%
\bibitem [{\citenamefont {Heeger}\ \emph {et~al.}(1988)\citenamefont {Heeger},
  \citenamefont {Kivelson}, \citenamefont {Schrieffer},\ and\ \citenamefont
  {Su}}]{Heeger_1988}%
  \BibitemOpen
  \bibfield  {author} {\bibinfo {author} {\bibfnamefont {A.~J.}\ \bibnamefont
  {Heeger}}, \bibinfo {author} {\bibfnamefont {S.}~\bibnamefont {Kivelson}},
  \bibinfo {author} {\bibfnamefont {J.~R.}\ \bibnamefont {Schrieffer}},\ and\
  \bibinfo {author} {\bibfnamefont {W.~P.}\ \bibnamefont {Su}},\ }\href
  {https://doi.org/10.1103/RevModPhys.60.781} {\bibfield  {journal} {\bibinfo
  {journal} {Rev. Mod. Phys.}\ }\textbf {\bibinfo {volume} {60}},\ \bibinfo
  {pages} {781} (\bibinfo {year} {1988})}\BibitemShut {NoStop}%
\bibitem [{\citenamefont {Fraxanet}\ \emph {et~al.}(2022)\citenamefont
  {Fraxanet}, \citenamefont {Gonz\'alez-Cuadra}, \citenamefont {Pfau},
  \citenamefont {Lewenstein}, \citenamefont {Langen},\ and\ \citenamefont
  {Barbiero}}]{fraxanet}%
  \BibitemOpen
  \bibfield  {author} {\bibinfo {author} {\bibfnamefont {J.}~\bibnamefont
  {Fraxanet}}, \bibinfo {author} {\bibfnamefont {D.}~\bibnamefont
  {Gonz\'alez-Cuadra}}, \bibinfo {author} {\bibfnamefont {T.}~\bibnamefont
  {Pfau}}, \bibinfo {author} {\bibfnamefont {M.}~\bibnamefont {Lewenstein}},
  \bibinfo {author} {\bibfnamefont {T.}~\bibnamefont {Langen}},\ and\ \bibinfo
  {author} {\bibfnamefont {L.}~\bibnamefont {Barbiero}},\ }\href
  {https://doi.org/10.1103/PhysRevLett.128.043402} {\bibfield  {journal}
  {\bibinfo  {journal} {Phys. Rev. Lett.}\ }\textbf {\bibinfo {volume} {128}},\
  \bibinfo {pages} {043402} (\bibinfo {year} {2022})}\BibitemShut {NoStop}%
\bibitem [{\citenamefont {Xing}\ \emph {et~al.}(2021)\citenamefont {Xing},
  \citenamefont {Chiu}, \citenamefont {Poletti}, \citenamefont {Scalettar},\
  and\ \citenamefont {Batrouni}}]{Xing2021}%
  \BibitemOpen
  \bibfield  {author} {\bibinfo {author} {\bibfnamefont {B.}~\bibnamefont
  {Xing}}, \bibinfo {author} {\bibfnamefont {W.-T.}\ \bibnamefont {Chiu}},
  \bibinfo {author} {\bibfnamefont {D.}~\bibnamefont {Poletti}}, \bibinfo
  {author} {\bibfnamefont {R.~T.}\ \bibnamefont {Scalettar}},\ and\ \bibinfo
  {author} {\bibfnamefont {G.}~\bibnamefont {Batrouni}},\ }\href
  {https://doi.org/10.1103/PhysRevLett.126.017601} {\bibfield  {journal}
  {\bibinfo  {journal} {Phys. Rev. Lett.}\ }\textbf {\bibinfo {volume} {126}},\
  \bibinfo {pages} {017601} (\bibinfo {year} {2021})}\BibitemShut {NoStop}%
\bibitem [{\citenamefont {Chanda}\ \emph {et~al.}(2021)\citenamefont {Chanda},
  \citenamefont {Kraus}, \citenamefont {Morigi},\ and\ \citenamefont
  {Zakrzewski}}]{chanda2021}%
  \BibitemOpen
  \bibfield  {author} {\bibinfo {author} {\bibfnamefont {T.}~\bibnamefont
  {Chanda}}, \bibinfo {author} {\bibfnamefont {R.}~\bibnamefont {Kraus}},
  \bibinfo {author} {\bibfnamefont {G.}~\bibnamefont {Morigi}},\ and\ \bibinfo
  {author} {\bibfnamefont {J.}~\bibnamefont {Zakrzewski}},\ }\href
  {https://doi.org/10.22331/q-2021-07-13-501} {\bibfield  {journal} {\bibinfo
  {journal} {{Quantum}}\ }\textbf {\bibinfo {volume} {5}},\ \bibinfo {pages}
  {501} (\bibinfo {year} {2021})}\BibitemShut {NoStop}%
\bibitem [{\citenamefont {Jaksch}\ and\ \citenamefont
  {Zoller}(2005)}]{Jaksch_2005}%
  \BibitemOpen
  \bibfield  {author} {\bibinfo {author} {\bibfnamefont {D.}~\bibnamefont
  {Jaksch}}\ and\ \bibinfo {author} {\bibfnamefont {P.}~\bibnamefont
  {Zoller}},\ }\href
  {https://doi.org/https://doi.org/10.1016/j.aop.2004.09.010} {\bibfield
  {journal} {\bibinfo  {journal} {Annals of Physics}\ }\textbf {\bibinfo
  {volume} {315}},\ \bibinfo {pages} {52} (\bibinfo {year} {2005})},\ \bibinfo
  {note} {special Issue}\BibitemShut {NoStop}%
\bibitem [{\citenamefont {Lewenstein}\ \emph {et~al.}(2007)\citenamefont
  {Lewenstein}, \citenamefont {Sanpera}, \citenamefont {Ahufinger},
  \citenamefont {Damski}, \citenamefont {Sen(De)},\ and\ \citenamefont
  {Sen}}]{Lewenstein_2007}%
  \BibitemOpen
  \bibfield  {author} {\bibinfo {author} {\bibfnamefont {M.}~\bibnamefont
  {Lewenstein}}, \bibinfo {author} {\bibfnamefont {A.}~\bibnamefont {Sanpera}},
  \bibinfo {author} {\bibfnamefont {V.}~\bibnamefont {Ahufinger}}, \bibinfo
  {author} {\bibfnamefont {B.}~\bibnamefont {Damski}}, \bibinfo {author}
  {\bibfnamefont {A.}~\bibnamefont {Sen(De)}},\ and\ \bibinfo {author}
  {\bibfnamefont {U.}~\bibnamefont {Sen}},\ }\href
  {https://doi.org/10.1080/00018730701223200} {\bibfield  {journal} {\bibinfo
  {journal} {Advances in Physics}\ }\textbf {\bibinfo {volume} {56}},\ \bibinfo
  {pages} {243} (\bibinfo {year} {2007})}\BibitemShut {NoStop}%
\bibitem [{\citenamefont {Gross}\ and\ \citenamefont
  {Bloch}(2017)}]{Gross_2017}%
  \BibitemOpen
  \bibfield  {author} {\bibinfo {author} {\bibfnamefont {C.}~\bibnamefont
  {Gross}}\ and\ \bibinfo {author} {\bibfnamefont {I.}~\bibnamefont {Bloch}},\
  }\href {https://doi.org/10.1126/science.aal3837} {\bibfield  {journal}
  {\bibinfo  {journal} {Science}\ }\textbf {\bibinfo {volume} {357}},\ \bibinfo
  {pages} {995} (\bibinfo {year} {2017})}\BibitemShut {NoStop}%
\bibitem [{\citenamefont {Gonz\'alez-Cuadra}\ \emph {et~al.}(2018)\citenamefont
  {Gonz\'alez-Cuadra}, \citenamefont {Grzybowski}, \citenamefont {Dauphin},\
  and\ \citenamefont {Lewenstein}}]{gonzalez2018}%
  \BibitemOpen
  \bibfield  {author} {\bibinfo {author} {\bibfnamefont {D.}~\bibnamefont
  {Gonz\'alez-Cuadra}}, \bibinfo {author} {\bibfnamefont {P.~R.}\ \bibnamefont
  {Grzybowski}}, \bibinfo {author} {\bibfnamefont {A.}~\bibnamefont
  {Dauphin}},\ and\ \bibinfo {author} {\bibfnamefont {M.}~\bibnamefont
  {Lewenstein}},\ }\href {https://doi.org/10.1103/PhysRevLett.121.090402}
  {\bibfield  {journal} {\bibinfo  {journal} {Phys. Rev. Lett.}\ }\textbf
  {\bibinfo {volume} {121}},\ \bibinfo {pages} {090402} (\bibinfo {year}
  {2018})}\BibitemShut {NoStop}%
\bibitem [{\citenamefont {Ritsch}\ \emph {et~al.}(2013)\citenamefont {Ritsch},
  \citenamefont {Domokos}, \citenamefont {Brennecke},\ and\ \citenamefont
  {Esslinger}}]{Ritsch_2013}%
  \BibitemOpen
  \bibfield  {author} {\bibinfo {author} {\bibfnamefont {H.}~\bibnamefont
  {Ritsch}}, \bibinfo {author} {\bibfnamefont {P.}~\bibnamefont {Domokos}},
  \bibinfo {author} {\bibfnamefont {F.}~\bibnamefont {Brennecke}},\ and\
  \bibinfo {author} {\bibfnamefont {T.}~\bibnamefont {Esslinger}},\ }\href
  {https://doi.org/10.1103/RevModPhys.85.553} {\bibfield  {journal} {\bibinfo
  {journal} {Rev. Mod. Phys.}\ }\textbf {\bibinfo {volume} {85}},\ \bibinfo
  {pages} {553} (\bibinfo {year} {2013})}\BibitemShut {NoStop}%
\bibitem [{\citenamefont {Mivehvar}\ \emph {et~al.}(2017)\citenamefont
  {Mivehvar}, \citenamefont {Ritsch},\ and\ \citenamefont
  {Piazza}}]{Mivehvar_2017}%
  \BibitemOpen
  \bibfield  {author} {\bibinfo {author} {\bibfnamefont {F.}~\bibnamefont
  {Mivehvar}}, \bibinfo {author} {\bibfnamefont {H.}~\bibnamefont {Ritsch}},\
  and\ \bibinfo {author} {\bibfnamefont {F.}~\bibnamefont {Piazza}},\ }\href
  {https://doi.org/10.1103/PhysRevLett.118.073602} {\bibfield  {journal}
  {\bibinfo  {journal} {Phys. Rev. Lett.}\ }\textbf {\bibinfo {volume} {118}},\
  \bibinfo {pages} {073602} (\bibinfo {year} {2017})}\BibitemShut {NoStop}%
\bibitem [{\citenamefont {Gonz\'alez-Cuadra}\ \emph {et~al.}(2019)\citenamefont
  {Gonz\'alez-Cuadra}, \citenamefont {Dauphin}, \citenamefont {Grzybowski},
  \citenamefont {W\'ojcik}, \citenamefont {Lewenstein},\ and\ \citenamefont
  {Bermudez}}]{gonzalez2019}%
  \BibitemOpen
  \bibfield  {author} {\bibinfo {author} {\bibfnamefont {D.}~\bibnamefont
  {Gonz\'alez-Cuadra}}, \bibinfo {author} {\bibfnamefont {A.}~\bibnamefont
  {Dauphin}}, \bibinfo {author} {\bibfnamefont {P.~R.}\ \bibnamefont
  {Grzybowski}}, \bibinfo {author} {\bibfnamefont {P.}~\bibnamefont
  {W\'ojcik}}, \bibinfo {author} {\bibfnamefont {M.}~\bibnamefont
  {Lewenstein}},\ and\ \bibinfo {author} {\bibfnamefont {A.}~\bibnamefont
  {Bermudez}},\ }\href {https://doi.org/10.1103/PhysRevB.99.045139} {\bibfield
  {journal} {\bibinfo  {journal} {Phys. Rev. B}\ }\textbf {\bibinfo {volume}
  {99}},\ \bibinfo {pages} {045139} (\bibinfo {year} {2019})}\BibitemShut
  {NoStop}%
\bibitem [{\citenamefont {Gonz{\'a}lez-Cuadra}\ \emph
  {et~al.}(2019)\citenamefont {Gonz{\'a}lez-Cuadra}, \citenamefont {Bermudez},
  \citenamefont {Grzybowski}, \citenamefont {Lewenstein},\ and\ \citenamefont
  {Dauphin}}]{gonzalez2019b}%
  \BibitemOpen
  \bibfield  {author} {\bibinfo {author} {\bibfnamefont {D.}~\bibnamefont
  {Gonz{\'a}lez-Cuadra}}, \bibinfo {author} {\bibfnamefont {A.}~\bibnamefont
  {Bermudez}}, \bibinfo {author} {\bibfnamefont {P.~R.}\ \bibnamefont
  {Grzybowski}}, \bibinfo {author} {\bibfnamefont {M.}~\bibnamefont
  {Lewenstein}},\ and\ \bibinfo {author} {\bibfnamefont {A.}~\bibnamefont
  {Dauphin}},\ }\href {https://doi.org/10.1038/s41467-019-10796-8} {\bibfield
  {journal} {\bibinfo  {journal} {Nature Communications}\ }\textbf {\bibinfo
  {volume} {10}},\ \bibinfo {pages} {2694} (\bibinfo {year}
  {2019})}\BibitemShut {NoStop}%
\bibitem [{\citenamefont {Chanda}\ \emph
  {et~al.}(2022{\natexlab{a}})\citenamefont {Chanda}, \citenamefont {Kraus},
  \citenamefont {Zakrzewski},\ and\ \citenamefont {Morigi}}]{chanda2022}%
  \BibitemOpen
  \bibfield  {author} {\bibinfo {author} {\bibfnamefont {T.}~\bibnamefont
  {Chanda}}, \bibinfo {author} {\bibfnamefont {R.}~\bibnamefont {Kraus}},
  \bibinfo {author} {\bibfnamefont {J.}~\bibnamefont {Zakrzewski}},\ and\
  \bibinfo {author} {\bibfnamefont {G.}~\bibnamefont {Morigi}},\ }\href
  {https://doi.org/10.1103/PhysRevB.106.075137} {\bibfield  {journal} {\bibinfo
   {journal} {Phys. Rev. B}\ }\textbf {\bibinfo {volume} {106}},\ \bibinfo
  {pages} {075137} (\bibinfo {year} {2022}{\natexlab{a}})}\BibitemShut
  {NoStop}%
\bibitem [{\citenamefont {Chanda}\ \emph
  {et~al.}(2022{\natexlab{b}})\citenamefont {Chanda}, \citenamefont
  {González-Cuadra}, \citenamefont {Lewenstein}, \citenamefont {Tagliacozzo},\
  and\ \citenamefont {Zakrzewski}}]{Chanda_2022}%
  \BibitemOpen
  \bibfield  {author} {\bibinfo {author} {\bibfnamefont {T.}~\bibnamefont
  {Chanda}}, \bibinfo {author} {\bibfnamefont {D.}~\bibnamefont
  {González-Cuadra}}, \bibinfo {author} {\bibfnamefont {M.}~\bibnamefont
  {Lewenstein}}, \bibinfo {author} {\bibfnamefont {L.}~\bibnamefont
  {Tagliacozzo}},\ and\ \bibinfo {author} {\bibfnamefont {J.}~\bibnamefont
  {Zakrzewski}},\ }\href {https://doi.org/10.21468/SciPostPhys.12.2.076}
  {\bibfield  {journal} {\bibinfo  {journal} {SciPost Phys.}\ }\textbf
  {\bibinfo {volume} {12}},\ \bibinfo {pages} {076} (\bibinfo {year}
  {2022}{\natexlab{b}})}\BibitemShut {NoStop}%
\bibitem [{\citenamefont {Manmana}\ \emph {et~al.}(2012)\citenamefont
  {Manmana}, \citenamefont {Essin}, \citenamefont {Noack},\ and\ \citenamefont
  {Gurarie}}]{manmana}%
  \BibitemOpen
  \bibfield  {author} {\bibinfo {author} {\bibfnamefont {S.~R.}\ \bibnamefont
  {Manmana}}, \bibinfo {author} {\bibfnamefont {A.~M.}\ \bibnamefont {Essin}},
  \bibinfo {author} {\bibfnamefont {R.~M.}\ \bibnamefont {Noack}},\ and\
  \bibinfo {author} {\bibfnamefont {V.}~\bibnamefont {Gurarie}},\ }\href
  {https://doi.org/10.1103/PhysRevB.86.205119} {\bibfield  {journal} {\bibinfo
  {journal} {Phys. Rev. B}\ }\textbf {\bibinfo {volume} {86}},\ \bibinfo
  {pages} {205119} (\bibinfo {year} {2012})}\BibitemShut {NoStop}%
\bibitem [{\citenamefont {Sirker}\ \emph {et~al.}(2014)\citenamefont {Sirker},
  \citenamefont {Maiti}, \citenamefont {Konstantinidis},\ and\ \citenamefont
  {Sedlmayr}}]{sirker}%
  \BibitemOpen
  \bibfield  {author} {\bibinfo {author} {\bibfnamefont {J.}~\bibnamefont
  {Sirker}}, \bibinfo {author} {\bibfnamefont {M.}~\bibnamefont {Maiti}},
  \bibinfo {author} {\bibfnamefont {N.~P.}\ \bibnamefont {Konstantinidis}},\
  and\ \bibinfo {author} {\bibfnamefont {N.}~\bibnamefont {Sedlmayr}},\ }\href
  {https://doi.org/10.1088/1742-5468/2014/10/p10032} {\bibfield  {journal}
  {\bibinfo  {journal} {Journal of Statistical Mechanics: Theory and
  Experiment}\ }\textbf {\bibinfo {volume} {2014}},\ \bibinfo {pages} {P10032}
  (\bibinfo {year} {2014})}\BibitemShut {NoStop}%
\bibitem [{\citenamefont {Wang}\ \emph {et~al.}(2015)\citenamefont {Wang},
  \citenamefont {Xu}, \citenamefont {Wang},\ and\ \citenamefont {Wu}}]{wang}%
  \BibitemOpen
  \bibfield  {author} {\bibinfo {author} {\bibfnamefont {D.}~\bibnamefont
  {Wang}}, \bibinfo {author} {\bibfnamefont {S.}~\bibnamefont {Xu}}, \bibinfo
  {author} {\bibfnamefont {Y.}~\bibnamefont {Wang}},\ and\ \bibinfo {author}
  {\bibfnamefont {C.}~\bibnamefont {Wu}},\ }\href
  {https://doi.org/10.1103/PhysRevB.91.115118} {\bibfield  {journal} {\bibinfo
  {journal} {Phys. Rev. B}\ }\textbf {\bibinfo {volume} {91}},\ \bibinfo
  {pages} {115118} (\bibinfo {year} {2015})}\BibitemShut {NoStop}%
\bibitem [{\citenamefont {Barbiero}\ \emph {et~al.}(2018)\citenamefont
  {Barbiero}, \citenamefont {Santos},\ and\ \citenamefont
  {Goldman}}]{barbiero2}%
  \BibitemOpen
  \bibfield  {author} {\bibinfo {author} {\bibfnamefont {L.}~\bibnamefont
  {Barbiero}}, \bibinfo {author} {\bibfnamefont {L.}~\bibnamefont {Santos}},\
  and\ \bibinfo {author} {\bibfnamefont {N.}~\bibnamefont {Goldman}},\ }\href
  {https://doi.org/10.1103/PhysRevB.97.201115} {\bibfield  {journal} {\bibinfo
  {journal} {Phys. Rev. B}\ }\textbf {\bibinfo {volume} {97}},\ \bibinfo
  {pages} {201115} (\bibinfo {year} {2018})}\BibitemShut {NoStop}%
\bibitem [{\citenamefont {Sbierski}\ and\ \citenamefont
  {Karrasch}(2018)}]{sbierski}%
  \BibitemOpen
  \bibfield  {author} {\bibinfo {author} {\bibfnamefont {B.}~\bibnamefont
  {Sbierski}}\ and\ \bibinfo {author} {\bibfnamefont {C.}~\bibnamefont
  {Karrasch}},\ }\href {https://doi.org/10.1103/PhysRevB.98.165101} {\bibfield
  {journal} {\bibinfo  {journal} {Phys. Rev. B}\ }\textbf {\bibinfo {volume}
  {98}},\ \bibinfo {pages} {165101} (\bibinfo {year} {2018})}\BibitemShut
  {NoStop}%
\bibitem [{\citenamefont {Yu}\ \emph {et~al.}(2020)\citenamefont {Yu},
  \citenamefont {Jiang}, \citenamefont {Quan}, \citenamefont {Wu},
  \citenamefont {Chen}, \citenamefont {Zou},\ and\ \citenamefont {Wu}}]{yu}%
  \BibitemOpen
  \bibfield  {author} {\bibinfo {author} {\bibfnamefont {X.-L.}\ \bibnamefont
  {Yu}}, \bibinfo {author} {\bibfnamefont {L.}~\bibnamefont {Jiang}}, \bibinfo
  {author} {\bibfnamefont {Y.-M.}\ \bibnamefont {Quan}}, \bibinfo {author}
  {\bibfnamefont {T.}~\bibnamefont {Wu}}, \bibinfo {author} {\bibfnamefont
  {Y.}~\bibnamefont {Chen}}, \bibinfo {author} {\bibfnamefont {L.-J.}\
  \bibnamefont {Zou}},\ and\ \bibinfo {author} {\bibfnamefont {J.}~\bibnamefont
  {Wu}},\ }\href {https://doi.org/10.1103/PhysRevB.101.045422} {\bibfield
  {journal} {\bibinfo  {journal} {Phys. Rev. B}\ }\textbf {\bibinfo {volume}
  {101}},\ \bibinfo {pages} {045422} (\bibinfo {year} {2020})}\BibitemShut
  {NoStop}%
\bibitem [{\citenamefont {Fraser}\ and\ \citenamefont
  {Piazza}(2019)}]{Fraser_2019}%
  \BibitemOpen
  \bibfield  {author} {\bibinfo {author} {\bibfnamefont {K.~A.}\ \bibnamefont
  {Fraser}}\ and\ \bibinfo {author} {\bibfnamefont {F.}~\bibnamefont
  {Piazza}},\ }\href {https://doi.org/10.1038/s42005-019-0149-1} {\bibfield
  {journal} {\bibinfo  {journal} {Communications Physics}\ }\textbf {\bibinfo
  {volume} {2}},\ \bibinfo {pages} {48} (\bibinfo {year} {2019})}\BibitemShut
  {NoStop}%
\bibitem [{\citenamefont {Gonz\'alez-Cuadra}\ \emph
  {et~al.}(2020{\natexlab{a}})\citenamefont {Gonz\'alez-Cuadra}, \citenamefont
  {Dauphin}, \citenamefont {Grzybowski}, \citenamefont {Lewenstein},\ and\
  \citenamefont {Bermudez}}]{Gonzalez-Cuadra_2020a}%
  \BibitemOpen
  \bibfield  {author} {\bibinfo {author} {\bibfnamefont {D.}~\bibnamefont
  {Gonz\'alez-Cuadra}}, \bibinfo {author} {\bibfnamefont {A.}~\bibnamefont
  {Dauphin}}, \bibinfo {author} {\bibfnamefont {P.~R.}\ \bibnamefont
  {Grzybowski}}, \bibinfo {author} {\bibfnamefont {M.}~\bibnamefont
  {Lewenstein}},\ and\ \bibinfo {author} {\bibfnamefont {A.}~\bibnamefont
  {Bermudez}},\ }\href {https://doi.org/10.1103/PhysRevLett.125.265301}
  {\bibfield  {journal} {\bibinfo  {journal} {Phys. Rev. Lett.}\ }\textbf
  {\bibinfo {volume} {125}},\ \bibinfo {pages} {265301} (\bibinfo {year}
  {2020}{\natexlab{a}})}\BibitemShut {NoStop}%
\bibitem [{\citenamefont {Gonz\'alez-Cuadra}\ \emph
  {et~al.}(2020{\natexlab{b}})\citenamefont {Gonz\'alez-Cuadra}, \citenamefont
  {Dauphin}, \citenamefont {Grzybowski}, \citenamefont {Lewenstein},\ and\
  \citenamefont {Bermudez}}]{Gonzalez-Cuadra_2020b}%
  \BibitemOpen
  \bibfield  {author} {\bibinfo {author} {\bibfnamefont {D.}~\bibnamefont
  {Gonz\'alez-Cuadra}}, \bibinfo {author} {\bibfnamefont {A.}~\bibnamefont
  {Dauphin}}, \bibinfo {author} {\bibfnamefont {P.~R.}\ \bibnamefont
  {Grzybowski}}, \bibinfo {author} {\bibfnamefont {M.}~\bibnamefont
  {Lewenstein}},\ and\ \bibinfo {author} {\bibfnamefont {A.}~\bibnamefont
  {Bermudez}},\ }\href {https://doi.org/10.1103/PhysRevB.102.245137} {\bibfield
   {journal} {\bibinfo  {journal} {Phys. Rev. B}\ }\textbf {\bibinfo {volume}
  {102}},\ \bibinfo {pages} {245137} (\bibinfo {year}
  {2020}{\natexlab{b}})}\BibitemShut {NoStop}%
\bibitem [{\citenamefont {Juli\`a-Farr\'e}\ \emph {et~al.}(2022)\citenamefont
  {Juli\`a-Farr\'e}, \citenamefont {Gonz\'alez-Cuadra}, \citenamefont
  {Patscheider}, \citenamefont {Mark}, \citenamefont {Ferlaino}, \citenamefont
  {Lewenstein}, \citenamefont {Barbiero},\ and\ \citenamefont
  {Dauphin}}]{Sergi2022}%
  \BibitemOpen
  \bibfield  {author} {\bibinfo {author} {\bibfnamefont {S.}~\bibnamefont
  {Juli\`a-Farr\'e}}, \bibinfo {author} {\bibfnamefont {D.}~\bibnamefont
  {Gonz\'alez-Cuadra}}, \bibinfo {author} {\bibfnamefont {A.}~\bibnamefont
  {Patscheider}}, \bibinfo {author} {\bibfnamefont {M.~J.}\ \bibnamefont
  {Mark}}, \bibinfo {author} {\bibfnamefont {F.}~\bibnamefont {Ferlaino}},
  \bibinfo {author} {\bibfnamefont {M.}~\bibnamefont {Lewenstein}}, \bibinfo
  {author} {\bibfnamefont {L.}~\bibnamefont {Barbiero}},\ and\ \bibinfo
  {author} {\bibfnamefont {A.}~\bibnamefont {Dauphin}},\ }\href
  {https://doi.org/10.1103/PhysRevResearch.4.L032005} {\bibfield  {journal}
  {\bibinfo  {journal} {Phys. Rev. Res.}\ }\textbf {\bibinfo {volume} {4}},\
  \bibinfo {pages} {L032005} (\bibinfo {year} {2022})}\BibitemShut {NoStop}%
\bibitem [{\citenamefont {Landig}\ \emph {et~al.}(2016)\citenamefont {Landig},
  \citenamefont {Hruby}, \citenamefont {Dogra}, \citenamefont {Landini},
  \citenamefont {Mottl}, \citenamefont {Donner},\ and\ \citenamefont
  {Esslinger}}]{Esslinger2016}%
  \BibitemOpen
  \bibfield  {author} {\bibinfo {author} {\bibfnamefont {R.}~\bibnamefont
  {Landig}}, \bibinfo {author} {\bibfnamefont {L.}~\bibnamefont {Hruby}},
  \bibinfo {author} {\bibfnamefont {N.}~\bibnamefont {Dogra}}, \bibinfo
  {author} {\bibfnamefont {M.}~\bibnamefont {Landini}}, \bibinfo {author}
  {\bibfnamefont {R.}~\bibnamefont {Mottl}}, \bibinfo {author} {\bibfnamefont
  {T.}~\bibnamefont {Donner}},\ and\ \bibinfo {author} {\bibfnamefont
  {T.}~\bibnamefont {Esslinger}},\ }\href {https://doi.org/10.1038/nature17409}
  {\bibfield  {journal} {\bibinfo  {journal} {Nature}\ }\textbf {\bibinfo
  {volume} {532}},\ \bibinfo {pages} {476} (\bibinfo {year}
  {2016})}\BibitemShut {NoStop}%
\bibitem [{\citenamefont {L{\'e}onard}\ \emph {et~al.}(2017)\citenamefont
  {L{\'e}onard}, \citenamefont {Morales}, \citenamefont {Zupancic},
  \citenamefont {Esslinger},\ and\ \citenamefont {Donner}}]{Leonard_2017}%
  \BibitemOpen
  \bibfield  {author} {\bibinfo {author} {\bibfnamefont {J.}~\bibnamefont
  {L{\'e}onard}}, \bibinfo {author} {\bibfnamefont {A.}~\bibnamefont
  {Morales}}, \bibinfo {author} {\bibfnamefont {P.}~\bibnamefont {Zupancic}},
  \bibinfo {author} {\bibfnamefont {T.}~\bibnamefont {Esslinger}},\ and\
  \bibinfo {author} {\bibfnamefont {T.}~\bibnamefont {Donner}},\ }\href
  {https://doi.org/10.1038/nature21067} {\bibfield  {journal} {\bibinfo
  {journal} {Nature}\ }\textbf {\bibinfo {volume} {543}},\ \bibinfo {pages}
  {87} (\bibinfo {year} {2017})}\BibitemShut {NoStop}%
\bibitem [{\citenamefont {Maschler}\ and\ \citenamefont
  {Ritsch}(2005)}]{maschler2005}%
  \BibitemOpen
  \bibfield  {author} {\bibinfo {author} {\bibfnamefont {C.}~\bibnamefont
  {Maschler}}\ and\ \bibinfo {author} {\bibfnamefont {H.}~\bibnamefont
  {Ritsch}},\ }\href {https://doi.org/10.1103/PhysRevLett.95.260401} {\bibfield
   {journal} {\bibinfo  {journal} {Phys. Rev. Lett.}\ }\textbf {\bibinfo
  {volume} {95}},\ \bibinfo {pages} {260401} (\bibinfo {year}
  {2005})}\BibitemShut {NoStop}%
\bibitem [{\citenamefont {Jaksch}\ \emph {et~al.}(1998)\citenamefont {Jaksch},
  \citenamefont {Bruder}, \citenamefont {Cirac}, \citenamefont {Gardiner},\
  and\ \citenamefont {Zoller}}]{Jaksch_1998}%
  \BibitemOpen
  \bibfield  {author} {\bibinfo {author} {\bibfnamefont {D.}~\bibnamefont
  {Jaksch}}, \bibinfo {author} {\bibfnamefont {C.}~\bibnamefont {Bruder}},
  \bibinfo {author} {\bibfnamefont {J.~I.}\ \bibnamefont {Cirac}}, \bibinfo
  {author} {\bibfnamefont {C.~W.}\ \bibnamefont {Gardiner}},\ and\ \bibinfo
  {author} {\bibfnamefont {P.}~\bibnamefont {Zoller}},\ }\href
  {https://doi.org/10.1103/PhysRevLett.81.3108} {\bibfield  {journal} {\bibinfo
   {journal} {Phys. Rev. Lett.}\ }\textbf {\bibinfo {volume} {81}},\ \bibinfo
  {pages} {3108} (\bibinfo {year} {1998})}\BibitemShut {NoStop}%
\bibitem [{SM()}]{SM}%
  \BibitemOpen
  \href@noop {} {}\bibinfo {note} {See Supplemental Material}\BibitemShut
  {NoStop}%
\bibitem [{\citenamefont {White}(1992)}]{dmrg}%
  \BibitemOpen
  \bibfield  {author} {\bibinfo {author} {\bibfnamefont {S.~R.}\ \bibnamefont
  {White}},\ }\href {https://doi.org/10.1103/PhysRevLett.69.2863} {\bibfield
  {journal} {\bibinfo  {journal} {Phys. Rev. Lett.}\ }\textbf {\bibinfo
  {volume} {69}},\ \bibinfo {pages} {2863} (\bibinfo {year}
  {1992})}\BibitemShut {NoStop}%
\bibitem [{\citenamefont {Hauschild}\ and\ \citenamefont
  {Pollmann}(2018)}]{tenpy}%
  \BibitemOpen
  \bibfield  {author} {\bibinfo {author} {\bibfnamefont {J.}~\bibnamefont
  {Hauschild}}\ and\ \bibinfo {author} {\bibfnamefont {F.}~\bibnamefont
  {Pollmann}},\ }\href {https://doi.org/10.21468/SciPostPhysLectNotes.5}
  {\bibfield  {journal} {\bibinfo  {journal} {SciPost Phys. Lect. Notes}\ ,\
  \bibinfo {pages} {5}} (\bibinfo {year} {2018})},\ \bibinfo {note} {code
  available from \url{https://github.com/tenpy/tenpy}},\ \Eprint
  {https://arxiv.org/abs/1805.00055} {arXiv:1805.00055} \BibitemShut {NoStop}%
\bibitem [{\citenamefont {Gong}\ \emph {et~al.}(2014)\citenamefont {Gong},
  \citenamefont {Zhu}, \citenamefont {Sheng}, \citenamefont {Motrunich},\ and\
  \citenamefont {Fisher}}]{shoushu2014}%
  \BibitemOpen
  \bibfield  {author} {\bibinfo {author} {\bibfnamefont {S.-S.}\ \bibnamefont
  {Gong}}, \bibinfo {author} {\bibfnamefont {W.}~\bibnamefont {Zhu}}, \bibinfo
  {author} {\bibfnamefont {D.~N.}\ \bibnamefont {Sheng}}, \bibinfo {author}
  {\bibfnamefont {O.~I.}\ \bibnamefont {Motrunich}},\ and\ \bibinfo {author}
  {\bibfnamefont {M.~P.~A.}\ \bibnamefont {Fisher}},\ }\href
  {https://doi.org/10.1103/PhysRevLett.113.027201} {\bibfield  {journal}
  {\bibinfo  {journal} {Phys. Rev. Lett.}\ }\textbf {\bibinfo {volume} {113}},\
  \bibinfo {pages} {027201} (\bibinfo {year} {2014})}\BibitemShut {NoStop}%
\bibitem [{\citenamefont {Wang}\ \emph {et~al.}(2016)\citenamefont {Wang},
  \citenamefont {Gu}, \citenamefont {Verstraete},\ and\ \citenamefont
  {Wen}}]{wangling2016}%
  \BibitemOpen
  \bibfield  {author} {\bibinfo {author} {\bibfnamefont {L.}~\bibnamefont
  {Wang}}, \bibinfo {author} {\bibfnamefont {Z.-C.}\ \bibnamefont {Gu}},
  \bibinfo {author} {\bibfnamefont {F.}~\bibnamefont {Verstraete}},\ and\
  \bibinfo {author} {\bibfnamefont {X.-G.}\ \bibnamefont {Wen}},\ }\href
  {https://doi.org/10.1103/PhysRevB.94.075143} {\bibfield  {journal} {\bibinfo
  {journal} {Phys. Rev. B}\ }\textbf {\bibinfo {volume} {94}},\ \bibinfo
  {pages} {075143} (\bibinfo {year} {2016})}\BibitemShut {NoStop}%
\bibitem [{\citenamefont {Mambrini}\ \emph {et~al.}(2006)\citenamefont
  {Mambrini}, \citenamefont {L\"auchli}, \citenamefont {Poilblanc},\ and\
  \citenamefont {Mila}}]{mambrini2006}%
  \BibitemOpen
  \bibfield  {author} {\bibinfo {author} {\bibfnamefont {M.}~\bibnamefont
  {Mambrini}}, \bibinfo {author} {\bibfnamefont {A.}~\bibnamefont {L\"auchli}},
  \bibinfo {author} {\bibfnamefont {D.}~\bibnamefont {Poilblanc}},\ and\
  \bibinfo {author} {\bibfnamefont {F.}~\bibnamefont {Mila}},\ }\href
  {https://doi.org/10.1103/PhysRevB.74.144422} {\bibfield  {journal} {\bibinfo
  {journal} {Phys. Rev. B}\ }\textbf {\bibinfo {volume} {74}},\ \bibinfo
  {pages} {144422} (\bibinfo {year} {2006})}\BibitemShut {NoStop}%
\bibitem [{\citenamefont {Pollmann}\ \emph {et~al.}(2010)\citenamefont
  {Pollmann}, \citenamefont {Turner}, \citenamefont {Berg},\ and\ \citenamefont
  {Oshikawa}}]{Pollmann2010}%
  \BibitemOpen
  \bibfield  {author} {\bibinfo {author} {\bibfnamefont {F.}~\bibnamefont
  {Pollmann}}, \bibinfo {author} {\bibfnamefont {A.~M.}\ \bibnamefont
  {Turner}}, \bibinfo {author} {\bibfnamefont {E.}~\bibnamefont {Berg}},\ and\
  \bibinfo {author} {\bibfnamefont {M.}~\bibnamefont {Oshikawa}},\ }\href
  {https://doi.org/10.1103/PhysRevB.81.064439} {\bibfield  {journal} {\bibinfo
  {journal} {Phys. Rev. B}\ }\textbf {\bibinfo {volume} {81}},\ \bibinfo
  {pages} {064439} (\bibinfo {year} {2010})}\BibitemShut {NoStop}%
\bibitem [{\citenamefont {Araki}\ \emph {et~al.}(2020)\citenamefont {Araki},
  \citenamefont {Mizoguchi},\ and\ \citenamefont {Hatsugai}}]{hatsugai2020}%
  \BibitemOpen
  \bibfield  {author} {\bibinfo {author} {\bibfnamefont {H.}~\bibnamefont
  {Araki}}, \bibinfo {author} {\bibfnamefont {T.}~\bibnamefont {Mizoguchi}},\
  and\ \bibinfo {author} {\bibfnamefont {Y.}~\bibnamefont {Hatsugai}},\ }\href
  {https://doi.org/10.1103/PhysRevResearch.2.012009} {\bibfield  {journal}
  {\bibinfo  {journal} {Phys. Rev. Res.}\ }\textbf {\bibinfo {volume} {2}},\
  \bibinfo {pages} {012009} (\bibinfo {year} {2020})}\BibitemShut {NoStop}%
\bibitem [{\citenamefont {Atala}\ \emph {et~al.}(2013)\citenamefont {Atala},
  \citenamefont {Aidelsburger}, \citenamefont {Barreiro}, \citenamefont
  {Abanin}, \citenamefont {Kitagawa}, \citenamefont {Demler},\ and\
  \citenamefont {Bloch}}]{Atala2013}%
  \BibitemOpen
  \bibfield  {author} {\bibinfo {author} {\bibfnamefont {M.}~\bibnamefont
  {Atala}}, \bibinfo {author} {\bibfnamefont {M.}~\bibnamefont {Aidelsburger}},
  \bibinfo {author} {\bibfnamefont {J.~T.}\ \bibnamefont {Barreiro}}, \bibinfo
  {author} {\bibfnamefont {D.}~\bibnamefont {Abanin}}, \bibinfo {author}
  {\bibfnamefont {T.}~\bibnamefont {Kitagawa}}, \bibinfo {author}
  {\bibfnamefont {E.}~\bibnamefont {Demler}},\ and\ \bibinfo {author}
  {\bibfnamefont {I.}~\bibnamefont {Bloch}},\ }\href
  {https://doi.org/10.1038/nphys2790} {\bibfield  {journal} {\bibinfo
  {journal} {Nature Physics}\ }\textbf {\bibinfo {volume} {9}},\ \bibinfo
  {pages} {795} (\bibinfo {year} {2013})}\BibitemShut {NoStop}%
\bibitem [{\citenamefont {Lohse}\ \emph {et~al.}(2016)\citenamefont {Lohse},
  \citenamefont {Schweizer}, \citenamefont {Zilberberg}, \citenamefont
  {Aidelsburger},\ and\ \citenamefont {Bloch}}]{Lohse_2016}%
  \BibitemOpen
  \bibfield  {author} {\bibinfo {author} {\bibfnamefont {M.}~\bibnamefont
  {Lohse}}, \bibinfo {author} {\bibfnamefont {C.}~\bibnamefont {Schweizer}},
  \bibinfo {author} {\bibfnamefont {O.}~\bibnamefont {Zilberberg}}, \bibinfo
  {author} {\bibfnamefont {M.}~\bibnamefont {Aidelsburger}},\ and\ \bibinfo
  {author} {\bibfnamefont {I.}~\bibnamefont {Bloch}},\ }\href
  {https://doi.org/10.1038/nphys3584} {\bibfield  {journal} {\bibinfo
  {journal} {Nature Physics}\ }\textbf {\bibinfo {volume} {12}},\ \bibinfo
  {pages} {350} (\bibinfo {year} {2016})}\BibitemShut {NoStop}%
\bibitem [{\citenamefont {Bakr}\ \emph {et~al.}(2009)\citenamefont {Bakr},
  \citenamefont {Gillen}, \citenamefont {Peng}, \citenamefont {F{\"o}lling},\
  and\ \citenamefont {Greiner}}]{Bakr_2009}%
  \BibitemOpen
  \bibfield  {author} {\bibinfo {author} {\bibfnamefont {W.~S.}\ \bibnamefont
  {Bakr}}, \bibinfo {author} {\bibfnamefont {J.~I.}\ \bibnamefont {Gillen}},
  \bibinfo {author} {\bibfnamefont {A.}~\bibnamefont {Peng}}, \bibinfo {author}
  {\bibfnamefont {S.}~\bibnamefont {F{\"o}lling}},\ and\ \bibinfo {author}
  {\bibfnamefont {M.}~\bibnamefont {Greiner}},\ }\href
  {https://doi.org/10.1038/nature08482} {\bibfield  {journal} {\bibinfo
  {journal} {Nature}\ }\textbf {\bibinfo {volume} {462}},\ \bibinfo {pages}
  {74} (\bibinfo {year} {2009})}\BibitemShut {NoStop}%
\bibitem [{\citenamefont {Sherson}\ \emph {et~al.}(2010)\citenamefont
  {Sherson}, \citenamefont {Weitenberg}, \citenamefont {Endres}, \citenamefont
  {Cheneau}, \citenamefont {Bloch},\ and\ \citenamefont {Kuhr}}]{Sherson_2010}%
  \BibitemOpen
  \bibfield  {author} {\bibinfo {author} {\bibfnamefont {J.~F.}\ \bibnamefont
  {Sherson}}, \bibinfo {author} {\bibfnamefont {C.}~\bibnamefont {Weitenberg}},
  \bibinfo {author} {\bibfnamefont {M.}~\bibnamefont {Endres}}, \bibinfo
  {author} {\bibfnamefont {M.}~\bibnamefont {Cheneau}}, \bibinfo {author}
  {\bibfnamefont {I.}~\bibnamefont {Bloch}},\ and\ \bibinfo {author}
  {\bibfnamefont {S.}~\bibnamefont {Kuhr}},\ }\href
  {https://doi.org/10.1038/nature09378} {\bibfield  {journal} {\bibinfo
  {journal} {Nature}\ }\textbf {\bibinfo {volume} {467}},\ \bibinfo {pages}
  {68} (\bibinfo {year} {2010})}\BibitemShut {NoStop}%
\bibitem [{\citenamefont {Kokail}\ \emph
  {et~al.}(2021{\natexlab{a}})\citenamefont {Kokail}, \citenamefont {van
  Bijnen}, \citenamefont {Elben}, \citenamefont {Vermersch},\ and\
  \citenamefont {Zoller}}]{Kokail_2021}%
  \BibitemOpen
  \bibfield  {author} {\bibinfo {author} {\bibfnamefont {C.}~\bibnamefont
  {Kokail}}, \bibinfo {author} {\bibfnamefont {R.}~\bibnamefont {van Bijnen}},
  \bibinfo {author} {\bibfnamefont {A.}~\bibnamefont {Elben}}, \bibinfo
  {author} {\bibfnamefont {B.}~\bibnamefont {Vermersch}},\ and\ \bibinfo
  {author} {\bibfnamefont {P.}~\bibnamefont {Zoller}},\ }\href
  {https://doi.org/10.1038/s41567-021-01260-w} {\bibfield  {journal} {\bibinfo
  {journal} {Nature Physics}\ }\textbf {\bibinfo {volume} {17}},\ \bibinfo
  {pages} {936} (\bibinfo {year} {2021}{\natexlab{a}})}\BibitemShut {NoStop}%
\bibitem [{\citenamefont {Kokail}\ \emph
  {et~al.}(2021{\natexlab{b}})\citenamefont {Kokail}, \citenamefont {Sundar},
  \citenamefont {Zache}, \citenamefont {Elben}, \citenamefont {Vermersch},
  \citenamefont {Dalmonte}, \citenamefont {van Bijnen},\ and\ \citenamefont
  {Zoller}}]{Kokail_2021_2}%
  \BibitemOpen
  \bibfield  {author} {\bibinfo {author} {\bibfnamefont {C.}~\bibnamefont
  {Kokail}}, \bibinfo {author} {\bibfnamefont {B.}~\bibnamefont {Sundar}},
  \bibinfo {author} {\bibfnamefont {T.~V.}\ \bibnamefont {Zache}}, \bibinfo
  {author} {\bibfnamefont {A.}~\bibnamefont {Elben}}, \bibinfo {author}
  {\bibfnamefont {B.}~\bibnamefont {Vermersch}}, \bibinfo {author}
  {\bibfnamefont {M.}~\bibnamefont {Dalmonte}}, \bibinfo {author}
  {\bibfnamefont {R.}~\bibnamefont {van Bijnen}},\ and\ \bibinfo {author}
  {\bibfnamefont {P.}~\bibnamefont {Zoller}},\ }\href
  {https://doi.org/10.1103/PhysRevLett.127.170501} {\bibfield  {journal}
  {\bibinfo  {journal} {Phys. Rev. Lett.}\ }\textbf {\bibinfo {volume} {127}},\
  \bibinfo {pages} {170501} (\bibinfo {year} {2021}{\natexlab{b}})}\BibitemShut
  {NoStop}%
\end{thebibliography}

\begin{thebibliography}{3}%
\makeatletter
\providecommand \@ifxundefined [1]{%
 \@ifx{#1\undefined}
}%
\providecommand \@ifnum [1]{%
 \ifnum #1\expandafter \@firstoftwo
 \else \expandafter \@secondoftwo
 \fi
}%
\providecommand \@ifx [1]{%
 \ifx #1\expandafter \@firstoftwo
 \else \expandafter \@secondoftwo
 \fi
}%
\providecommand \natexlab [1]{#1}%
\providecommand \enquote  [1]{``#1''}%
\providecommand \bibnamefont  [1]{#1}%
\providecommand \bibfnamefont [1]{#1}%
\providecommand \citenamefont [1]{#1}%
\providecommand \href@noop [0]{\@secondoftwo}%
\providecommand \href [0]{\begingroup \@sanitize@url \@href}%
\providecommand \@href[1]{\@@startlink{#1}\@@href}%
\providecommand \@@href[1]{\endgroup#1\@@endlink}%
\providecommand \@sanitize@url [0]{\catcode `\\12\catcode `\$12\catcode
  `\&12\catcode `\#12\catcode `\^12\catcode `\_12\catcode `\%12\relax}%
\providecommand \@@startlink[1]{}%
\providecommand \@@endlink[0]{}%
\providecommand \url  [0]{\begingroup\@sanitize@url \@url }%
\providecommand \@url [1]{\endgroup\@href {#1}{\urlprefix }}%
\providecommand \urlprefix  [0]{URL }%
\providecommand \Eprint [0]{\href }%
\providecommand \doibase [0]{https://doi.org/}%
\providecommand \selectlanguage [0]{\@gobble}%
\providecommand \bibinfo  [0]{\@secondoftwo}%
\providecommand \bibfield  [0]{\@secondoftwo}%
\providecommand \translation [1]{[#1]}%
\providecommand \BibitemOpen [0]{}%
\providecommand \bibitemStop [0]{}%
\providecommand \bibitemNoStop [0]{.\EOS\space}%
\providecommand \EOS [0]{\spacefactor3000\relax}%
\providecommand \BibitemShut  [1]{\csname bibitem#1\endcsname}%
\let\auto@bib@innerbib\@empty
\bibitem [{\citenamefont {Maschler}\ and\ \citenamefont
  {Ritsch}(2005)}]{maschler2005}%
  \BibitemOpen
  \bibfield  {author} {\bibinfo {author} {\bibfnamefont {C.}~\bibnamefont
  {Maschler}}\ and\ \bibinfo {author} {\bibfnamefont {H.}~\bibnamefont
  {Ritsch}},\ }\href {https://doi.org/10.1103/PhysRevLett.95.260401} {\bibfield
   {journal} {\bibinfo  {journal} {Phys. Rev. Lett.}\ }\textbf {\bibinfo
  {volume} {95}},\ \bibinfo {pages} {260401} (\bibinfo {year}
  {2005})}\BibitemShut {NoStop}%
\bibitem [{\citenamefont {Chanda}\ \emph {et~al.}(2021)\citenamefont {Chanda},
  \citenamefont {Kraus}, \citenamefont {Morigi},\ and\ \citenamefont
  {Zakrzewski}}]{chanda2021}%
  \BibitemOpen
  \bibfield  {author} {\bibinfo {author} {\bibfnamefont {T.}~\bibnamefont
  {Chanda}}, \bibinfo {author} {\bibfnamefont {R.}~\bibnamefont {Kraus}},
  \bibinfo {author} {\bibfnamefont {G.}~\bibnamefont {Morigi}},\ and\ \bibinfo
  {author} {\bibfnamefont {J.}~\bibnamefont {Zakrzewski}},\ }\href
  {https://doi.org/10.22331/q-2021-07-13-501} {\bibfield  {journal} {\bibinfo
  {journal} {{Quantum}}\ }\textbf {\bibinfo {volume} {5}},\ \bibinfo {pages}
  {501} (\bibinfo {year} {2021})}\BibitemShut {NoStop}%
\bibitem [{\citenamefont {Ritsch}\ \emph {et~al.}(2013)\citenamefont {Ritsch},
  \citenamefont {Domokos}, \citenamefont {Brennecke},\ and\ \citenamefont
  {Esslinger}}]{Ritsch_2013}%
  \BibitemOpen
  \bibfield  {author} {\bibinfo {author} {\bibfnamefont {H.}~\bibnamefont
  {Ritsch}}, \bibinfo {author} {\bibfnamefont {P.}~\bibnamefont {Domokos}},
  \bibinfo {author} {\bibfnamefont {F.}~\bibnamefont {Brennecke}},\ and\
  \bibinfo {author} {\bibfnamefont {T.}~\bibnamefont {Esslinger}},\ }\href
  {https://doi.org/10.1103/RevModPhys.85.553} {\bibfield  {journal} {\bibinfo
  {journal} {Rev. Mod. Phys.}\ }\textbf {\bibinfo {volume} {85}},\ \bibinfo
  {pages} {553} (\bibinfo {year} {2013})}\BibitemShut {NoStop}%
\end{thebibliography}
\end{document}